\documentclass[aps,prd,superscriptaddress,amsmath,groupedaddress,twocolumn]{revtex4}
\usepackage{color,hangcaption,hhline,psfrag,rotating,amssymb}
\usepackage[hang,nooneline]{subfigure}
\usepackage{dcolumn}

\begin{document}
\title{Update: Precision $D_s$ decay constant from full lattice QCD using very fine lattices}

\author{C. T. H. Davies}
\email[]{c.davies@physics.gla.ac.uk}
\affiliation{Department of Physics and Astronomy, University of Glasgow, Glasgow, G12 8QQ, UK}
\author{C. McNeile}
\thanks{Current address: Dept. of Theoretical Physics, Bergische Universit\"{a}t Wuppertal, Gaussstr.\,20, D-42119 Wuppertal, Germany  }
\affiliation{Department of Physics and Astronomy, University of Glasgow, Glasgow, G12 8QQ, UK}
\author{E. Follana}
\affiliation{Departamento de F\'{\i}sica Te\'{o}rica, Universidad de Zaragoza, E-50009 Zaragoza, Spain}
\author{G. P. Lepage}
\affiliation{Laboratory of Elementary-Particle Physics, Cornell University, Ithaca, New York 14853, USA}
\author{H. Na}
\affiliation{Department of Physics, The Ohio State University, Columbus, Ohio, 43210, USA}
\author{J. Shigemitsu}
\affiliation{Department of Physics, The Ohio State University, Columbus, Ohio, 43210, USA}
%\thanks{}
%\affiliation{}

\collaboration{HPQCD collaboration}
\homepage{http://www.physics.gla.ac.uk/HPQCD}
\noaffiliation

\date{\today}

\begin{abstract}
We update our previous determination of both the decay constant and the 
mass of the $D_s$ meson using the Highly Improved Staggered Quark 
formalism. We include additional results at two finer values
of the lattice spacing along with improved determinations of the 
lattice spacing and improved tuning of the charm and strange quark masses. 
We obtain $m_{D_s}$ = 1.9691(32) GeV, in good agreement with experiment, 
 and $f_{D_s}$ = 0.2480(25) GeV. Our result for $f_{D_s}$ is 
1.6$\sigma$ lower than the most recent experimental 
average determined from the $D_s$ leptonic decay rate and using 
$V_{cs}$ from CKM unitarity. 
Combining our $f_{D_s}$ with the experimental rate we obtain 
a direct determination of $V_{cs} = 1.010(22)$, or 
alternatively $0.990 {+0.013 \atop -0.016}$ using a probability distribution for statistical errors for this 
quantity which vanishes above 1.  
We also include an accurate prediction of the decay constant of the $\eta_c$, 
$f_{\eta_c}$ = 0.3947(24) GeV, as a calibration point for other lattice calculations. 
\end{abstract}

%\pacs{}
%\keywords{}

\maketitle

\section{Introduction}

Lattice QCD is now a firmly established method for providing precision tests 
of the Standard Model~\cite{ourlatqcd}. Combined with experiment, lattice QCD calculations
have the potential to uncover new physics provided that both the 
theoretical and experimental results are accurate enough. 

The most accurate lattice QCD calculations are those for the masses of 
`gold-plated' mesons, where few MeV errors are now possible across 
the entire spectrum. This accuracy is at the level where electromagnetic 
effects on the meson masses, currently missing from lattice 
QCD calculations, have to be estimated and included. 
Ref.~\cite{bcstar} gives a recent summary including 
predictions of masses that have been made ahead of experiment.
The meson masses are extracted from simple `two-point' hadron correlation 
functions calculated on the lattice from combining 
appropriate valence quark and antiquark propagators. 
Another parallel set of quantities that can be determined from the same correlation 
functions are the meson decay constants. Calculations of 
these can be compared to experimental results for rates
of annihilation to photons for neutral unflavored vector mesons and
to $W$ bosons for charged pseudoscalars. 
By determining as complete and accurate a picture as possible for decay constants 
along with masses we provide a stringent test of the Standard Model. 
Physics beyond the Standard Model can introduce new ways to 
decay to leptons for some mesons, and so accurate comparison of decay constants 
between theory and experiment can also provide direct constraints on new physics 
models. 

Here we focus on results for one quantity, the decay constant of the 
$D_s$ meson, $f_{D_s}$, which has been a showcase for the impact that 
accurate lattice QCD calculations can have, particularly when ahead of experimental results. 
We will update our result from 2007~\cite{fds}, making several improvements 
to the calculation. It is important to understand that $f_{D_s}$ is not calculated 
in isolation; as discussed above, it is one piece of the range of QCD physics that is calculable on 
the lattice. The other pieces, where they can also be tested against experiment, 
lend weight to the confidence we have in our error analysis. This is particularly 
true for our calculation because we can calculate a range of different quantities 
all with the same method. So here we also update 
our results for the mass of the $D_s$ meson and discuss other calculations
that will provide further tests. 
First we review briefly some background to the calculation of $f_{D_s}$. 

Decay constants for light pseudoscalar mesons ($f_{\pi}$ and $f_K$) have been calculable 
with errors at the few percent level since 2004~\cite{milc2004}, being one of 
the first calculations done in lattice QCD once ensembles of gluon field configurations were available 
that included the full effect of $u$, $d$ and $s$ sea quarks with a light enough 
mass for the $u/d$ quark to enable controlled extrapolation to the physical point.  
These calculations were done using the improved staggered (asqtad) formalism~\cite{gplasqtad, orginos} 
which has a number of advantages over previous formalisms, that 
mean that the calculation of $f_{\pi}$ and $f_K$ can be done accurately. 
Key requirements for these calculations are a quark formalism (such as improved staggered quarks) which : 
\begin{itemize}
\item has an 
absolutely normalised operator to couple to the $W$ boson; 
\item is improved so that it has small 
discretisation errors ($\cal{O}$$(\alpha_s a^2)$ for improved staggered quarks) and
\item is numerically 
fast so that large ensembles of gluon field configurations can be made 
including sea quarks and so that
many hadron correlation functions can be calculated per configuration, for small statistical errors. 
In addition a large physical volume ($> 2.5 {\rm fm}^3$) is necessary so that finite volume effects are reduced to the 1\% level. Having all of these features 
means that accurate extrapolations to the physical continuum limit can 
be made. 
\end{itemize}

Calculations of decay constants for mesons containing the heavier $c$ quark became 
important with the promise of results from the CLEO-c experiment. The first lattice results for $f_D$ and 
$f_{D_s}$ appeared  from the Fermilab Lattice/MILC collaborations 
in 2005 as predictions ahead of experiment~\cite{fnalfd}. 
They used the `Fermilab' formalism~\cite{fermilab}, 
developed many years previously for heavy quark physics,
and had errors of 8\%. This led to the unfortunate impression that decay constants 
for $D$ and $D_s$ mesons were inevitably much less accurate than those for 
$\pi$ and $K$ and errors would only be slowly reduced as higher statistics and the 
advent of finer 
lattices reduced statistical errors and systematic errors from discretisation 
effects. Because the Fermilab formalism predated the improved staggered 
formalism, however, these calculations had not made use of any of the features 
discussed above that made $f_{\pi}$ and $f_K$ so accurate.

For $c$ quarks the issue of discretisation errors becomes more 
important than for the lighter quarks. In 2007 we showed that further 
improving the improved staggered formalism to the Highly Improved 
Staggered (HISQ) formalism~\cite{hisq} produces a quark formalism that 
has all the good features of the asqtad formalism outlined 
above but also significantly smaller discretisation errors. 
In fact the discretisation errors are small enough that HISQ can be 
used for $c$ quarks as well as $u/d$ and $s$ quarks and using 
the same formalism for all 4 lightest quarks has enormous advantages.
We used HISQ for all the valence quarks
to calculate all 4 decay constants: $f_{\pi}$, $f_K$, $f_D$ and 
$f_{D_s}$ to better than 2\% accuracy~\cite{fds}. Our results were:
\begin{eqnarray}
f_{\pi} &=& 132(2) {\rm MeV} \\ \nonumber 
f_K &=& 157(2) {\rm MeV} \\ \nonumber
f_D &=& 207(4) {\rm MeV} \\ \nonumber
f_{D_s} &=& 241(3) {\rm MeV}
\end{eqnarray}
Although $f_D$ and 
$f_{D_s}$ still have noticeably larger discretisation errors (and 
therefore contributions to the systematic error from the extrapolation 
to the $a \rightarrow 0$ limit) than 
$f_K$ and $f_{\pi}$ there are smaller systematic errors from, for example, finite 
volume effects. This leads then to the expectation, and the result, of
very similar final errors. 
Our error for $f_{D_s}$ was somewhat smaller than that for $f_D$ (1.3\% versus 1.8\%) 
since the $D_s$ contains no valence $u/d$ quarks and is therefore much 
less sensitive to the chiral extrapolation to the physical $u/d$ quark mass. 
This makes $f_{D_s}$ a particularly accurate quantity to calculate in lattice QCD. 

Since, at that time, $f_D$ and $f_{D_s}$ 
were only known to 6-8\% from experiment~\cite{cleo07fd, cleo07fds, babar07fds}, we had the added test, unavailable 
to the Fermilab formalism, of agreeing with experiment for $f_{\pi}$ and $f_K$. 
An additional very stringent test that had not previously been done was the 
determination of the mass of the $D_s$ and $D$ mesons along with their decay 
constants. The masses are known to better than 1 MeV experimentally. We were 
able to achieve errors from lattice QCD of 7 MeV (0.3\%) by determining the 
difference between the $D$ or $D_s$ mass and one half that of the 
$\eta_c$. Electromagnetic effects on the masses, 
missing from the lattice QCD calculation, had to be allowed for in 
achieving this accuracy. Good agreement between lattice QCD and the 
experimental results was obtained. We quoted $m_{D_s}$ = 1.962(6) GeV 
and $m_D$ = 1.868(7) GeV~\cite{fds}. 

Following our result much improved experimental 
numbers for $f_D$~\cite{cleofd} (206(9) MeV) and 
$f_{D_s}$ (274(11) MeV) became available from CLEO~\cite{cleofds08}.
This produced the exciting picture in the summer of 2008 that agreement between experiment 
and our result for $f_D$ was very good but that the experimental result for 
$f_{D_s}$ (including averages with results from BaBar and Belle~\cite{bellefds}) 
was significantly larger than our lattice QCD value, see, for example,~\cite{cleo-ichep08, me-ichep08}. 
Since the experimental errors were still much bigger than 
ours the discrepancy, of $3\sigma$, was dominated by the 
experimental error. A burst of activity from other lattice QCD calculations produced 
results that agreed with 
ours but, having errors several times larger, often also agreed with the experimental one~\cite{proclat08}. 
This led to much speculation about the existence of new physics (that had to affect 
$D_s$ but not $D$)~\cite{kronfeld} as well as limits on new physics from 
the fact that the experimental $f_{D_s}$ was larger (and not smaller) than the Standard Model 
result from our calculation~\cite{akeroyd}.  

Since then improved statistics and further results from other 
channels~\cite{cleofdsmu, cleofdste, cleofdstrho, babar10} have brought down the experimental average and reduced 
its error to 2\%.  In early 2010, the Heavy Flavor Averaging Group (HFAG) 
gave a world average result from 
experiment of $f_{D_s}$ = 0.2546(59) GeV~\cite{hfag}, 
$2\sigma$ above our 2007 result. 
In the meantime we have extended our lattice QCD calculation 
using HISQ quarks to even finer lattices as well as improving the 
accuracy with which we determine the lattice spacing (which provides 
the calibration of the energy scale) and fix
the $c$ and $s$ quark masses. 
This has led our result for $f_{D_s}$ to move upwards, as we show here, to 
0.2480(25) GeV, with a slight improvement in the error to 1\%. 
The main reason for the upward shift 
is the recalibration of the lattice spacing. 
The experimental average as of October 2010 has moved up 
again slightly with new results from BaBar~\cite{babar10new}. 
Our value for $f_{D_s}$ is now 1.6$\sigma$ 
from the experimental average and this 
reduces considerably the room for new physics in this quantity. 

In section~\ref{sec:latt} we describe the lattice QCD calculation
and in section~\ref{sec:res} the results. These sections contain 
technical details which may not be of interest to those without 
a lattice QCD background. As well as $f_{D_s}$ we give results for 
$m_{D_s}$ which, as discussed above, is an important check on the 
calculation. We also show results for $f_{\eta_c}$, the decay constant 
of the $\eta_c$. This cannot be accessed directly from experiment but 
provides an excellent `figure of merit' for lattice QCD calculations in 
charm physics. We give the result to 0.6\% so that other lattice QCD 
calculations can compare to this when quoting numbers for $f_{D_s}$. 
In section~\ref{sec:disc} we discuss 
the picture that emerges from the current experimental and lattice 
QCD results for $f_{D_s}$, including the update we give here. 
We have tried to make this section readable by those that skipped 
the earlier technical details. We will also comment on the effects of 
the recalibration of the lattice energy scale on 
the other calculations included in~\cite{fds}, i.e. 
$f_K$, $f_{\pi}$ and $f_D$.  
Section~\ref{sec:conc} gives our conclusions.

\section{Lattice QCD calculation}
\label{sec:latt}

We work with 11 different ensembles of gluon field configurations 
provided by the MILC collaboration. These include the effect of 
$u$, $d$ and $s$ sea quarks using the improved staggered (asqtad) 
formalism and the fourth root `trick'. This procedure has passed 
various tests indicating that it is a valid discretization of QCD~\cite{eigs1, eigs2, staggtest}.
Configurations are available with large spatial volumes ($> 2.4 {\rm fm}^3$) 
for a wide range of values of the lattice spacing, $a$, and 
at multiple values of the sea light and strange quark masses.
The $u$ and $d$ quark masses are taken to be equal in the sea ($m_u = m_d = m_l$) for numerical speed. This has negligible effect on the calculations 
described here. 
We use configurations at 5 values of the lattice spacing
between 0.15 fm and 0.045 fm with parameters as listed in Table~\ref{tab:params}.
We have chosen the ensembles so that we can test the dependence of 
our results on each of: the lattice spacing; the physical volume; 
the sea light quark mass and the 
sea strange quark mass. 

\begin{table}
\caption{\label{tab:params}
Ensembles (sets) of MILC configurations with 
size $L^3 \times T$ and sea 
mass parameters $m_{l}^\mathrm{asq}$ and $m_{s}^\mathrm{asq}$ used for this analysis. 
The sea 
ASQTAD quark masses ($l = u/d$) are given in the MILC convention where $u_0$ is the plaquette 
tadpole parameter. Values of $u_0$ are given in Table~\ref{tab:delta}. 
The lattice spacing values in units of $r_1$ after `smoothing'
are given in the second column~\cite{milcreview}. 
Sets 1 and 2 are `very coarse'; sets 3, 4, 5, 6 and 7 `coarse'; sets 8 and 9
`fine'; set 10 `superfine' and set 11 `ultrafine'. The final column gives the 
number of configurations and the number of time sources per configuration 
used for calculating quark propagators for the best-tuned parameter 
sets on each ensemble. }
\begin{ruledtabular}
\begin{tabular}{lllllll}
Set & $r_1/a$ & $au_0m_{l}^\mathrm{asq}$ & $au_0m_{s}^\mathrm{asq}$ & $L/T$ & $N_{cf}\times N_t$ \\
\hline
1 &  2.152(5) & 0.0097 & 0.0484 & 16/48 & $631 \times 2$\\
2 &  2.138(4) & 0.0194 & 0.0484 & 16/48 & $631 \times 2$\\
\hline
3 &  2.647(3) & 0.005 & 0.05 & 24/64  & $678 \times 2$ \\
4 &  2.618(3) & 0.01 & 0.05 & 20/64 & $595 \times2$\\
5 &  2.618(3) & 0.01 & 0.05 & 28/64 & $269 \times 4$ \\
6 &  2.644(3) & 0.02 & 0.05 & 20/64 & $600 \times 2$ \\
7 &  2.658(3) & 0.01 & 0.03 & 20/64 & $328 \times 2$ \\
\hline
8 &  3.699(3) & 0.0062 & 0.031 & 28/96 & $566 \times 4$ \\
9 &  3.712(4) & 0.0124 & 0.031 & 28/96 & $600 \times 4$ \\
\hline 
10 &  5.296(7) & 0.0036 & 0.018 & 48/144 & $201 \times 2$ \\
\hline
11 &  7.115(20) & 0.0028 & 0.014 & 64/192 & $208 \times 1$\\
\end{tabular}
\end{ruledtabular}
\end{table}

On these configurations we have calculated quark propagators 
for charm quarks and strange quarks using the HISQ action. 
The numerical speed of HISQ means that we have been able to use 
several nearby quark masses for charm and strange to allow 
accurate interpolation to the correct values. This is 
described in the next section. These propagators are combined 
together to make pseudoscalar meson correlators with valence 
quark content either `charm-charm', `charm-strange' or 
`strange-strange'. By fitting the correlators as a function of 
the time separation of the source and the sink on the lattice 
we are able to determine the ground-state pseudoscalar mass 
(i.e. that of the $\eta_c$, $D_s$ or $\eta_s$) and the 
amplitude with which the ground-state meson is created 
or destroyed by the local temporal axial current. This 
latter quantity is directly related to the decay constant.  

The HISQ action~\cite{hisq} is an extension of the asqtad improved staggered 
quark action, which is itself based on the unimproved (naive) 
staggered quark action. The unimproved staggered action is 
equivalent to a simple `naive' discretization of the continuum 
quark action to give, on the lattice: 
\begin{equation}
S = \sum_x \overline{\psi}(x) \left( \gamma \cdot \Delta(U) + ma \right) \psi(x).
\end{equation}  
where $ma$ is the quark mass in lattice units. 
$\Delta(U)$ is a discrete version of the covariant derivative coupling to the 
lattice gluon field $U_{\mu}(x)$, which is a set of SU(3) matrices sitting on 
the links of the lattice:
\begin{equation}
\Delta_{\mu}(U)\psi(x) = \frac{1}{2} \left[ U_{\mu}(x)\psi(x+\hat{\mu}) - U_{\mu}^{\dagger}(x-\hat{\mu})\psi(x-\hat{\mu})\right].
\label{eq:deriv}
\end{equation}
In the improved staggered formalism the gluon field in the 
covariant derivative is smeared i.e. $U_{\mu}(x)$ is replaced 
by a sum of products of $U_{\mu}$ matrices tracing out more 
complicated paths between $x$ and $x+\hat{\mu}$~\cite{gplasqtad}. 
The smearing introduces a form factor that reduces the coupling 
between the quark and high momentum ( $p \approx \pi/a$) gluons 
that cause a particular type of discretisation error for staggered 
quarks. This error in principle appears at $\alpha_s a^2$ but in 
practice is very large for unimproved staggered quarks. The error 
is seen most clearly in the mass differences between different 
`tastes' of pseudoscalar meson, created by different point-split 
pseudoscalar operators. These mass splittings are proportional 
to $a^2$ and are strongly reduced on going from unimproved staggered 
quarks to improved staggered quarks~\cite{orginos}. 
Most smearing methods introduce additional discretisation errors. This 
is avoided here by the specific form of the smearing used~\cite{gplasqtad}. 
In the highly improved staggered quark action this smearing is applied 
twice with a reunitarisation of the gluon field in between. We also 
apply a projection back on to SU(3) for the gluon field, although this 
makes little difference in practice. 
We then find another further large reduction in the splittings between 
different tastes of pseudoscalar mesons~\cite{hisq}. 
In the pseudoscalar case the splitting in the squared 
masses ($\Delta m_{\pi}^2$) is roughly constant (for quark masses that are not too large) 
and so the splittings in the pseudoscalar masses themselves ($\Delta m_{\pi}$) fall 
with quark mass. Thus these `taste-changing' errors are generally
smaller for charm quarks than strange quarks~\cite{hisq, hisq2}, and they 
are particularly small with the HISQ action. 

Other, more mundane, discretisation errors are tackled by standard 
improvement techniques. 
A simple analysis in Fourier space of the symmetric difference of equation~\ref{eq:deriv} shows 
that this has errors of $\cal{O}$$(a^2)$ which 
can be corrected by the addition of a $(pa)^3$ term. This term, known as the Naik term~\cite{naik}, 
appears in the improved staggered quark action as a mixture of 3-link and 1-link 
differences. The improved staggered quark action then has discretisation errors 
that are $\cal{O}$$(a^4)$, apart from radiatively generated errors at $\cal{O}$$(\alpha_sa^2)$. 
The HISQ action uses the same Naik term (except that 
it contains smeared gluon fields) but corrects it further for discretisation 
errors when using quark masses appropriate to charm or heavier. 
Discretisation errors controlled by the quark mass $ma$ become important 
in that case. 
and we adjust the 
coefficient of the Naik term so that it takes value $(1+\epsilon)$ instead of 
1~\cite{hisq}. Then, schematically,  
\begin{equation}
S = \sum_x \overline{\psi}(x) \left( \gamma \cdot \tilde{\Delta}(U) + ma \right) \psi(x).
\end{equation}  
where
\begin{equation}
\tilde{\Delta}_{\mu} = \Delta_{\mu} - \frac{1+\epsilon}{6} \Delta^3_{\mu}.
\end{equation}
$\epsilon$ is a function of $ma$ (starting at $(ma)^2$) calculated to give the 
correct quark dispersion relation (energy as a function of momentum) 
at tree level. Here we give an exact formula for $\epsilon$ 
at tree level, $\epsilon_{tree}$, 
given an expansion for the tree level pole mass, $m_{tree}$, as a 
function of the mass $ma$ in the lattice action~\cite{hisq}: 
\begin{eqnarray}
m_{tree}a &=& ma [1 - \frac{3}{80}(ma)^4 + \frac{23}{2240}(ma)^6 \\ \nonumber
&&  \frac{1783}{537600}(ma)^8 - \frac{76943}{23654400}(ma)^{10} + \ldots ],
\end{eqnarray}
\begin{eqnarray}
\epsilon_{tree} + 1 = \frac{4 - \sqrt{4 + \frac{12m_{tree}a}{\cosh(m_{tree}a)\sinh(m_{tree}a)}}}{(\sinh(m_{tree}a))^2}.
\end{eqnarray}
These equations are obtained by solving the condition for the `kinetic mass', 
$M_2 = [\partial^2 E/\partial p_x^2]^{-1}$, to be equal to the 
tree level pole mass, $m_{tree}$. $m_{tree}$ in turn solves the 
pole condition at zero momentum.  
Including a Naik coefficient of $(1+\epsilon_{tree})$ means that 
the leading (in the velocity expansion) $(ma)^4$ errors are removed in the HISQ case, 
and so remaining discretisation errors are suppressed either by 
$\alpha_s$ or by the fact that heavy quarks are nonrelativistic in their 
bound states. 
$\epsilon$ can be fixed nonperturbatively by demanding that the `speed of 
light' be 1, and this was done in earlier 
calculations~\cite{hisq}. However it was found that nonperturbative results 
for $\epsilon$ were close to the tree level result in the HISQ case 
and so here we simply 
define $\epsilon$ to take the value $\epsilon_{tree}$ above.  

It is numerically very fast to calculate quark propagators for staggered 
actions because they have only one spin component. This means that we 
can readily calculate propagators from several different time sources on 
the lattice for improved statistics. Table~\ref{tab:params} lists the number 
of configurations used from each ensemble and the number of time sources 
per configuration. To increase statistics further we use a `random wall' 
source for the quark propagator instead of a delta function~\cite{milc2004}. The random wall 
is a set of U(1) random numbers with unit norm on every point of the source 
time slice (separately for each color) 
and is used as the source for the inversion to calculate the 
quark propagator. The same random wall is used for all propagators from a 
given time source on a given configuration so that when any 
propagator is combined with the complex conjugate of another
 to form a meson correlator 
the random numbers cancel except where the initial spatial points and 
colors are the same. This effectively increases the number of meson correlators 
sampled and reduces the statistical noise by a large factor for the case 
of pseudoscalar mesons. 
We also take a random starting point for our time sources for the very coarse, coarse and fine 
ensembles. 

The pseudoscalar meson correlation function $C_{ab}(t)$ for meson of 
valence content $a\overline{b}$ is calculated by multiplying
together the quark propagator for quark $a$ and the complex conjugate of 
the quark propagator for quark $b$ from the same source on a given 
configuration, 
matching colors at the source and
sink and matching the sink spatial site index, which is summed over 
to set the meson to zero momentum. 
The meson correlation function is then averaged over time sources 
on a single configuration. This means that any correlations between 
the time sources on a given configuration are accounted for. 
We also have to worry about autocorrelations between results on 
successive configurations in an ensemble. Tests by binning correlators 
have shown that the results on different configurations are independent 
of each other except on the finest lattices. We therefore bin the 
correlators on superfine and ultrafine lattices by a factor of two. 

The correlation function averaged over the independent samples 
from an ensemble is then fit as a function of the 
time separation between source and sink, $t$, to the form: 
\begin{equation}
\overline{C}(t) = \sum_i a_i (e^{-M_i t} + e^{-M_i (T-t)})
\label{eq:fit1}
\end{equation}
for the case $a=b$. $i=0$ is the ground state and larger $i$ values 
denote radial or other excitations with the same $J^{PC}$ quantum 
numbers. $T$ is the time extent of the lattice. 
For the unequal mass case there are additional 
`oscillating' terms coming from opposite parity states, denoted $i_p$: 
\begin{equation}
\overline{C}(t) = \sum_{i,i_p} a_i e^{-M_i t} + (-1)^ta_{i_p}e^{-M_{i_p} t} + ( t \rightarrow T- t)
\label{eq:fit2}
\end{equation}

To fit we use a number of exponentials $i$, and where appropriate $i_p$, 
in the range 2--6, loosely constraining the higher order exponentials 
by the use of Bayesian priors~\cite{gplbayes}. As the number of exponentials increases, 
we see the $\chi^2$ value fall below 1 and the results
for the fitted values 
and their errors for the parameters for the ground state $i=0$ stabilise. 
This allows us to determine the ground state parameters $a_0$ and $M_0$ as accurately 
as possible whilst allowing the full systematic error from the presence 
of higher excitations in the correlation function. 
We take the fit parameters 
to be the logarithm of the ground state masses $M_0$ and $M_{0_p}$ 
and the logarithms of 
the differences in mass between successive radial excitations (which 
are then forced to be positive). 
The Bayesian prior value for $M_0$ is obtained from a simple `effective 
mass' in the correlator and the prior width on the value is taken as a 
factor of 1.5. 
The prior value for the mass splitting between higher excitations 
is taken as roughly 600 MeV with a width of a factor of 2. 
Where oscillating states appear in the fit, 
the prior value for $M_{0_p}$
is taken as roughly 600 MeV above $M_0$ with a 
prior width of a factor of 2 and the 
splitting between higher oscillating excitations is taken to be 
the same as for the non-oscillating states. 
The amplitudes $a_i$ and $a_{i_p}$ are given prior 
widths of 1.0. 

Our fit includes the effect of correlation between 
different values of $t$.
We apply a cut on the range of eigenvalues from the 
correlation matrix that are used in the fit of $10^{-3}$ 
or $10^{-4}$. We also cut out very small $t$ values from our 
fit, 
typically below 3 or 4, to reduce the effect of higher 
excitations. 

The results for masses and amplitudes from fits in equations~\ref{eq:fit1} 
and~\ref{eq:fit2}
are in units of the lattice spacing.
The value of the lattice spacing must be determined for 
each ensemble to enable conversion 
to physical units. 
For this we use the parameter $r_1$, defined from the heavy quark 
potential~\cite{milc2004}. This parameter can be accurately determined (to better than 
0.5\%) in 
units of the lattice spacing and so is good for making ensemble to 
ensemble comparisons of $a$. Results from the MILC collaboration 
are given in Table~\ref{tab:params}. Unfortunately $r_1$ does not have a 
directly accessible physical value. Instead we must determine 
that from other quantities. In~\cite{r1paper} we used four other 
physical quantities with well-known experimental values to fix 
the value of $r_1$ to 0.3133(23) fm. 
This then yields a value for $a$ on a given ensemble with two 
errors - an error from the value of $r_1/a$ on that ensemble 
and an error, correlated between ensembles,
from the physical value of $r_1$. 

The amplitude, $a_0$, from the fits in equations~\ref{eq:fit1} and~\ref{eq:fit2}
is directly related to the matrix element for the local 
pseudoscalar operator to create or destroy the ground-state pseudoscalar meson 
from the vacuum. From the definition of the correlator and using 
a relativistic normalisation for the fields: 
\begin{equation}
a_0 = (\langle 0 | P_s | P \rangle )^2/2M_0
\label{eq:adef}
\end{equation}
where the pseudoscalar current $P_s = \overline{a}\gamma_5 b$ for 
quark content $a\overline{b}$. 
Because of the chiral symmetry of the staggered 
quark action we have a partially conserved axial current (PCAC) relation 
that relates the local pseudoscalar operator above to a temporal 
axial current that is absolutely normalised on the lattice. 
This allows us to determine the decay constant for these 
pseudoscalar mesons without having to worry about an uncertainty 
from the renormalisation between the lattice and the continuum. 
The decay constant for meson $P$ with quark content $a\overline{b}$ is defined from: 
\begin{equation}
\langle 0 | \overline{a} \gamma_{\mu} \gamma_5 b | P(p) \rangle \equiv f_{P} p_{\mu}.
\label{eq:fdef}
\end{equation}
For a meson at zero momentum, 
and using the PCAC relation $\partial_{\mu}A^{\mu} = (m_a + m_b) P_s$ 
to relate the axial vector 
and pseudoscalar currents, this becomes:
\begin{equation}
(m_a + m_b) \langle 0 | \overline{a} \gamma_5 b | P(p) \rangle \equiv f_{P} M_P^2,
\label{eq:fdef2}
\end{equation}
where $m_a$ and $m_b$ are the appropriate quark masses. 
Combining this with equation~\ref{eq:adef} then allows 
us to determine $f_P$ in lattice QCD from our fits to the correlators 
for pseudoscalar meson $P$ using 
\begin{equation}
f_P = (m_a + m_b) \sqrt{\frac{2a_0}{M_0^3}}.
\end{equation}
Here $m_a$ and $m_b$ are the quark masses used in 
the lattice QCD calculation. 

$f_P$ in turn is related, for charged pseudoscalars such as the $\pi$, $K$, $D$ and $D_s$ mesons, 
to the experimentally measurable leptonic branching fraction via a $W$ boson:
\begin{equation}
{\cal{B}}(P \rightarrow l \nu_l (\gamma)) = \frac{G_F^2 |V_{ab}|^2\tau_P}{8\pi}f_{P}^2m_l^2m_{P}\left( 1-\frac{m_l^2}{m_{P}^2}\right)^2,
\label{eq:gamma}
\end{equation}
up to calculable electromagnetic corrections. $V_{ab}$ is the appropriate CKM element for quark content $a\overline{b}$. $\tau_P$ is the pseudoscalar meson lifetime. 

\section{Results}
\label{sec:res}

Accurate results for the $D_s$ meson require accurate tuning 
of both the $c$ and the $s$ quark masses. We use the pseudoscalar 
mesons made purely of $c$ quarks or of $s$ quarks to do this and 
so first discuss results for these mesons. 

Table~\ref{tab:charmmass} lists the valence HISQ quark masses
close to that of the charm quark that we used for each of the 
gluon configuration ensembles along with the corresponding value of the 
Naik parameter $(1+\epsilon)$.  We also list the values of 
the ground-state pseudoscalar $c\overline{c}$ meson mass and 
decay constant obtained from our fits to the $c\overline{c}$ 
meson correlators to equation~\ref{eq:fit1}. The decay constant, $f_{\eta_c}$, 
will be discussed in subsection~\ref{sec:fetac} - it is a useful 
quantity to calculate despite the fact that the $\eta_c$ is a neutral 
particle and does not undergo a purely leptonic decay of the kind 
given in equation~\ref{eq:gamma}. To tune the charm 
quark mass in the HISQ action we must interpolate to the point at which 
the mass of the $\eta_c$ has the correct physical value on each ensemble. 
This physical value is not exactly the experimental value (2.980 GeV~\cite{pdg}) 
because our lattice QCD calculation corresponds to a world in which there
are no electromagnetic interactions and we do not allow our 
$\eta_c$ meson to annihilate to gluons. Both of these effects tend to 
reduce the $\eta_c$ mass by small amounts and so the appropriate
physical value for us to compare our lattice QCD calculation to is 
2.985(3) GeV, allowing a 50\% error for each correction
to the experimental value. The corrections are obtained from a potential 
model for the electromagnetic effect and from perturbation 
theory for the effect of gluon annihilation~\cite{hisq,r1paper}. 

\begin{table*}
\begin{ruledtabular}
\begin{tabular}{lllllll}
Set & $am_{c}$ &   $1+\epsilon$ & $am_{\eta_c}$ & $af_{\eta_c}$ & $am_{s}$ & $am_{\eta_s}$ \\
\hline
1 & 0.81 & 0.665 & 2.19381(16) & 0.3491(5) & 0.061 & 0.50490(36) \\
 & 0.825 &  0.656 & 2.22013(15) & 0.3539(5)  & 0.066 & 0.52524(36) \\ 
 & 0.85 &  0.641 & 2.26352(15) & 0.3622(5) & 0.080 & 0.57828(34) \\
2 & 0.825 &  0.656 & 2.21954(13) & 0.3537(4) & 0.066 & 0.52458(35) \\
\hline
3 & 0.622 & 0.779 & 1.79132(8) & 0.25706(18) & 0.0489 & 0.41133(17) \\ 
  & 0.65 &  0.762 & 1.84578(8) & 0.26368(18) & 0.0537 & 0.43118(18) \\
4 & 0.63 & 0.774 & 1.80849(11) & 0.25998(20) & 0.0492 & 0.41436(23) \\
 & 0.66 &  0.756 & 1.86666(10) & 0.26721(20) & 0.0546 & 0.43654(24) \\
 & 0.72 &  0.720 & 1.98109(10) & 0.28228(22) & 0.06 & 0.45787(23) \\
 & 0.753 &  0.700 & 2.04293(10) & 0.29114(24) & 0.063 & 0.46937(24) \\
5 & 0.63 & 0.774 & 1.80856(7) & 0.26006(15) & 0.0492 & 0.41457(14) \\
6 & 0.625 & 0.777 &  1.79347(13) & 0.2556(3) & 0.0491 & 0.41196(24) \\
 & & & & & 0.0525 & 0.42588(30) \\
& & & & & 0.0556 & 0.43834(30) \\
7 & 0.619 & 0.781 & 1.78595(15) & 0.2564(3) & 0.0487 & 0.41030(31) \\
\hline
8 & 0.413 & 0.893 & 1.28057(7) & 0.17217(11) & 0.0337 & 0.29413(12) \\
  & 0.43 &  0.885 & 1.31691(7) & 0.17508(11) & 0.0358 & 0.30332(12) \\
  & 0.44 &  0.880 & 1.33816(7) & 0.17678(11) & 0.0366 & 0.30675(12) \\  
  & 0.45 &  0.875 & 1.35934(7) & 0.17850(11) & 0.0382 & 0.31362(14) \\
9  & 0.412 & 0.894 & 1.27522(7) & 0.17086(10) & 0.0336 & 0.29309(13) \\
 & 0.427 &  0.885 & 1.30731(10) & 0.17344(15) & 0.03635 & 0.30513(20) \\
\hline
10 & 0.273 & 0.951  & 0.89935(12) & 0.11864(24) & 0.0228 & 0.20621(19) \\
  & 0.28 & 0.949 & 0.91543(8) & 0.11986(21) & 0.024 & 0.21196(13) \\
\hline
11 & 0.193 & 0.975 & 0.66628(13) & 0.0882(3) & 0.0161 & 0.15278(28) \\ 
 & 0.195 & 0.975 & 0.67117(6) & 0.08846(11) & 0.0165 & 0.15484(14) \\
  &  & &  &  & 0.018 & 0.16209(17) \\
\end{tabular}
\end{ruledtabular}
\caption{ Results for the masses in lattice units of the goldstone pseudoscalars made 
from valence HISQ charm or strange quarks on the different 
MILC ensembles, enumerated in Table~\ref{tab:params}. Columns 2 and 3 give 
the corresponding bare charm quark mass, 
and Naik coefficient respectively. Column 6 gives the bare strange 
quark mass ($\epsilon=0$ in that case). A lot of the meson masses in 
this table appear also in~\cite{mcspaper}
but we have added results on the coarse 02/05 and 01/03 ensembles (sets 6 and 7) and the large volume 
coarse 01/05 ensemble (set 5) as well as improving the tuning of masses 
on other ensembles and improving some fits on sets 4, 10 and 11. 
Results for the decay constant of the $\eta_c$ meson are also included, for 
analysis in subsection~\ref{sec:fetac}. 
 }
\label{tab:charmmass}
\end{table*}

\begin{figure}
\begin{center}
\includegraphics[width=80mm]{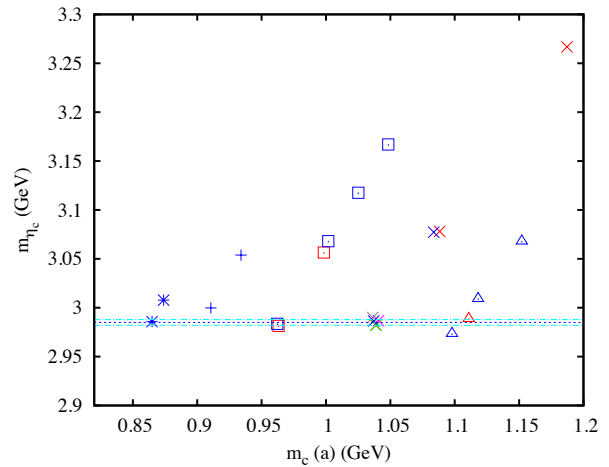}
\end{center}
\caption{Results for the mass of the pseudoscalar meson made of 
quarks with masses close to that of the charm quark mass for the 
full set of ensembles from Table~\ref{tab:params}. 
The $x$-axis is the lattice bare mass of the quark, which 
runs with lattice spacing from right to left. 
Very coarse 
ensembles are triangles; coarse, crosses; fine, squares; superfine, pluses; 
ultrafine, bursts. 
Results for heavier sea $u/d$ quark masses at 
each lattice spacing are in 
red, lighter ones are in blue. 
On the coarse lattices the very heavy sea masses of set 6 
are in pink, the lighter strange sea mass of set 7 in grey 
and the large volume results on set 5 are in green, 
on top of the result from set 4. 
Statistical errors are too small to be visible on this plot. 
The results show that tuning the quark mass to that of charm 
depends very little on the sea quark masses or on the volume. 
The dotted line gives the physical value, with its error,
 appropriate to lattice QCD, see text.}
\label{fig:fmhhsea}
\end{figure}

Figure~\ref{fig:fmhhsea} shows the meson mass in physical units plotted 
against the quark mass, also in physical units, for each ensemble. 
This plot demonstrates how the quark mass tuning is done, as well 
as illustrating very clearly how accurately we can do this from lattice 
QCD. Several features of the figure stand out. On a given ensemble 
the value of the meson mass is linear in the quark mass, as we 
would expect. The lines showing this behaviour (not plotted 
on the figure) are essentially parallel with a 
slope close to the naive expectation of 2 for 
ensembles with different lattice spacing values. 
In fact the slope does increase from 1.7 on the very coarse lattices 
to 2.3 on the superfine lattices. The reason for this is that 
the $x$-axis is a well-defined `running' quark mass, being the 
quark mass in the HISQ Lagrangian with a particular ultraviolet 
scale set by the lattice spacing. This is why we denote the 
mass on the $x$-axis as $m_c(a)$. 
The horizontal line indicates the correct value of the 
$\eta_c$ mass and therefore, where it cuts each set of results, 
the tuned value of $m_c$ at that lattice spacing. These values 
`run' to the left on finer lattices as the ultraviolet cut-off 
increases, as expected from perturbation 
theory. We expect the variation of $\eta_c$ mass with quark 
mass to be some number (say, 2) times the quark mass at a fixed 
scale. Therefore on finer lattices, where the scale is higher, 
we expect the slope to be larger, as demonstrated in Figure~\ref{fig:fmhhsea}. 

\begin{figure}
\begin{center}
\includegraphics[width=80mm]{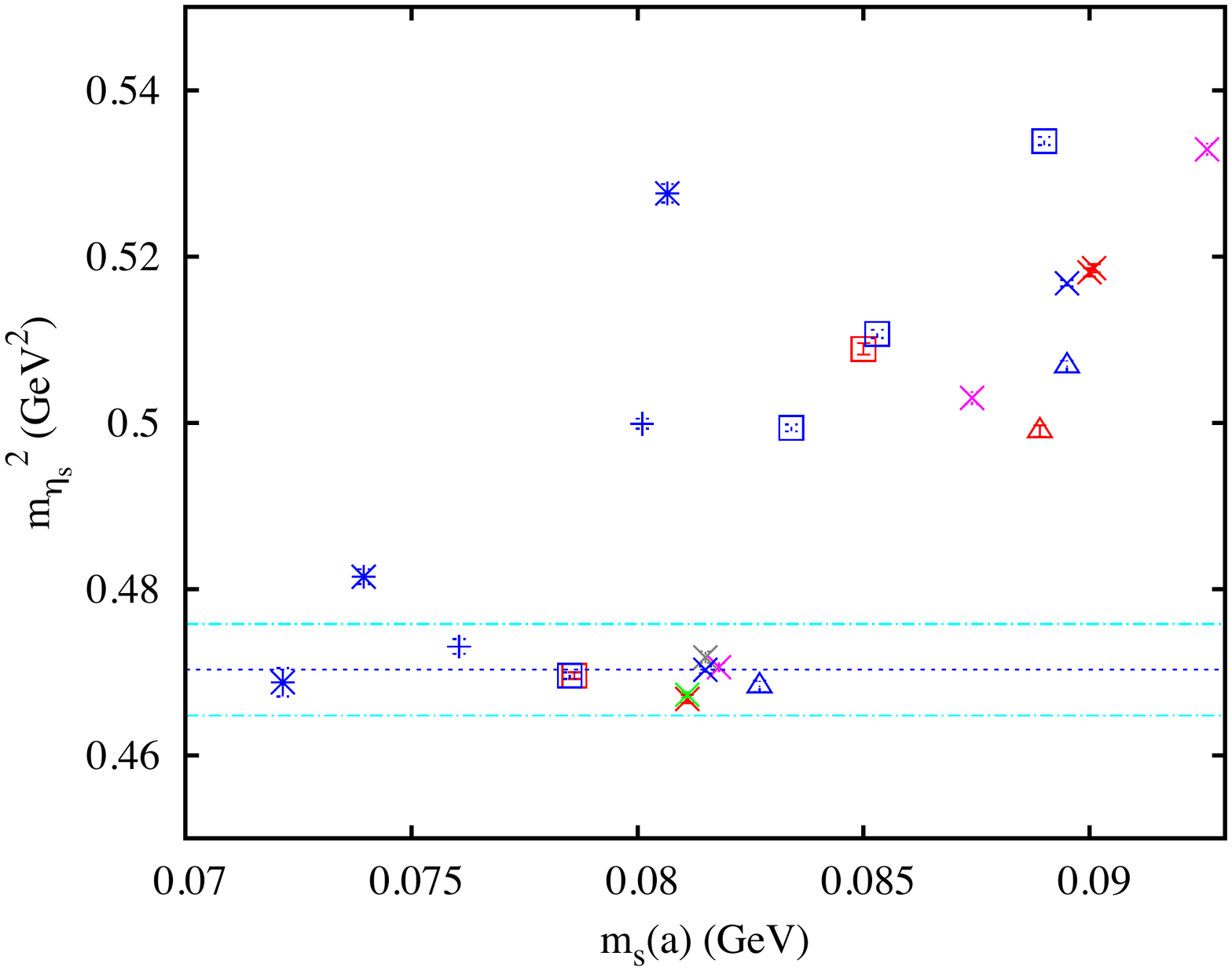}
\end{center}
\caption{Results for the square of the mass of the pseudoscalar meson 
made of 
quarks with masses close to that of the strange quark mass for the 
full set of ensembles from Table~\ref{tab:params}. 
Errors are statistical errors from the fits to the meson 
correlators. 
The $x$-axis is the lattice bare mass of the quark, which 
runs with lattice spacing from right to left. 
Very coarse 
ensembles are triangles; coarse, crosses; fine, squares; superfine, pluses; 
ultrafine, bursts. 
Results for heavier sea $u/d$ quark masses at 
each lattice spacing are in 
red, lighter ones are in blue. 
On the coarse lattices the very heavy sea masses of set 6 
are in pink, the lighter strange sea mass of set 7 in grey 
and the large volume results on set 5 are in green, 
on top of the result from set 4. 
The results show that tuning the quark mass to that of strange 
depends very little on the sea quark masses or on the volume.
The dotted line gives the physical value, with its error,
 appropriate to lattice QCD, see text.}
\label{fig:fmsssea}
\end{figure}

Another feature is that the results for different ensembles with very similar 
values of the lattice spacing are very close together
i.e. there is very little dependence of the tuned $c$ mass 
on the sea quark masses. 
The results for different physical volumes (sets 4 and 5) lie 
on top of each other showing that there is no dependence on the 
volume. We would not expect any significant volume dependence 
on these large spatial volumes
for the $\eta_c$ since it is a relatively small particle. 

From the horizontal line on Figure~\ref{fig:fmhhsea} and the lattice 
points on the line it is clear that we have tuned the charm quark 
mass very well on all except the superfine lattices (where it is off by  
0.1\%). In each case this corresponds to the lightest charm quark 
mass in our Table~\ref{tab:charmmass}. 
Figure~\ref{fig:fmhhsea} does not include errors in converting the 
lattice quark mass or $\eta_c$ mass to GeV coming from the 
values of $r_1/a$ or the physical value of $r_1$. The effect of 
these errors is reduced over naive expectations because 
$\Delta m_{\eta_c}$ is close to $2 \Delta m_c$, and so the leading 
order change from any $\Delta a$ cancels out. 
This issue was addressed in~\cite{mcspaper}. Here we are not aiming 
to determine $m_c$, but simply to make sure we understand the 
errors in other quantities induced by the tuning error in $m_ca$, 
so we leave a more detailed discussion of this source of systematic 
error to the sections 
on the individual quantities. 

Table~\ref{tab:charmmass} lists the valence HISQ quark masses
close to that of the strange quark that we used for making 
strange quark propagators on each of the gluon field ensembles. 
We also list the corresponding values of the mass of the 
ground-state $s\overline{s}$ meson, the $\eta_s$. The $\eta_s$ 
is not a particle available to study in the real world where 
$s\overline{s}$ mixes with $u\overline{u}$ and $d\overline{d}$. 
However, by omitting these possibilities in the lattice QCD 
calculation we can obtain a pure $s\overline{s}$ `pion-like' meson. 
This turns out to be useful for tuning the $s$ quark mass because 
the $\eta_s$ mass can be determined relatively precisely, and is 
less sensitive to the sea quark masses than, for example, $m_{\pi}$. 
However, the physical value for the $\eta_s$ mass has to be 
determined by relating it to $\pi$ and $K$ meson masses known 
from experiment. In earlier lattice QCD calculations we determined 
$m_{\eta_s}$ = 0.6858(40) GeV~\cite{r1paper} and this is the value we will use here.  
We also studied the $\eta_s$ decay constant which is again a quantity 
that cannot be measured experimentally in the real world but one
which turns out to be useful for determining the lattice spacing. 
We will not discuss $f_{\eta_s}$ further here. 

Figure~\ref{fig:fmsssea} shows the square of the $\eta_s$ mass against 
the quark mass, both in physical units, for each ensemble. We expect 
$m_{\eta_s}^2 \propto m_s$ from leading order chiral perturbation theory 
and the results indeed show this dependence. 
Once again the lines demonstrating this (not plotted on the 
figure) are fairly parallel but with a slope 
increasing on the finer lattices as the quark mass for a given meson 
mass runs to smaller values. The horizontal plots gives the physical 
value of the $\eta_s$ mass given above and the strange quark mass can be read 
off for each ensemble from where this crosses the line of data. Again 
we have well-tuned strange quark masses at each value of the lattice 
spacing at the lightest end of the range. The strange quark mass values 
on the very coarse and coarse lattices are rather close together but 
on the finer lattices the strange mass changes as rapidly with 
lattice spacing as the charm mass does in Figure~\ref{fig:fmhhsea}. 
In the continuum limit the ratio of these two masses becomes a
scale-invariant constant~\cite{mcspaper}. 

Again it is evident from Figure~\ref{fig:fmsssea} that 
there is very little dependence of the tuned $s$ quark mass 
on either the sea quark masses or the volume. 
Because the value of the tuned $s$ quark mass is proportional 
to the square of the $\eta_s$ mass the relative uncertainty in 
$m_s$ arising 
from lattice spacing errors is equal to that of the lattice 
spacing. There is no cancellation as there was in the case of 
the charm quark. In addition the 0.6\% uncertainty in the physical 
value of the $\eta_s$ mass is significant, because it becomes an 
uncertainty of 1.2\% in $m_s$. 
The effect of these uncertainties on the mass and decay constant 
of the $D_s$ meson will be discussed below. 

The staggered quarks in the sea are asqtad improved staggered 
quarks rather than HISQ quarks, i.e. they use a different 
discretisation of the quark piece of the QCD Lagrangian. 
The $s$ quark mass in the two formalisms will then not be the 
same, and we need to understand the ratio of the two so 
that we can extrapolate to the physical (real world) point 
for both the valence and sea quark masses.  We can determine
the physical points for the sea quark masses from our 
tuning of the valence masses and this ratio. 
There is very little sea quark mass dependence in the 
quantities that we study here, so that we do not need to 
know this ratio accurately. It is discussed further in 
Appendix~\ref{app:sea}. 

Once we have determined the $c$ and $s$ masses to be 
used to give the required physical results for 
the $\eta_c$ and the $\eta_s$ mesons, the $D_s$ 
meson correlator is entirely prescribed. There are 
no further adjustable parameters, given the nature
of QCD. The fit to the $D_s$ meson correlators 
gives us both the $D_s$ meson mass (from $M_0$ in 
equation~\ref{eq:fit2}) and its decay 
constant (from $a_0$) as testable outputs from 
lattice QCD. Since the $D_s$ meson mass is well-known 
experimentally it provides an excellent independent 
test of the error analysis on the decay constant. 
It is therefore very important to analyse both of 
these quantities together. 

\subsection{$m_{D_s}$}
\label{sec:mds}

\begin{figure}
\begin{center}
\includegraphics[width=80mm]{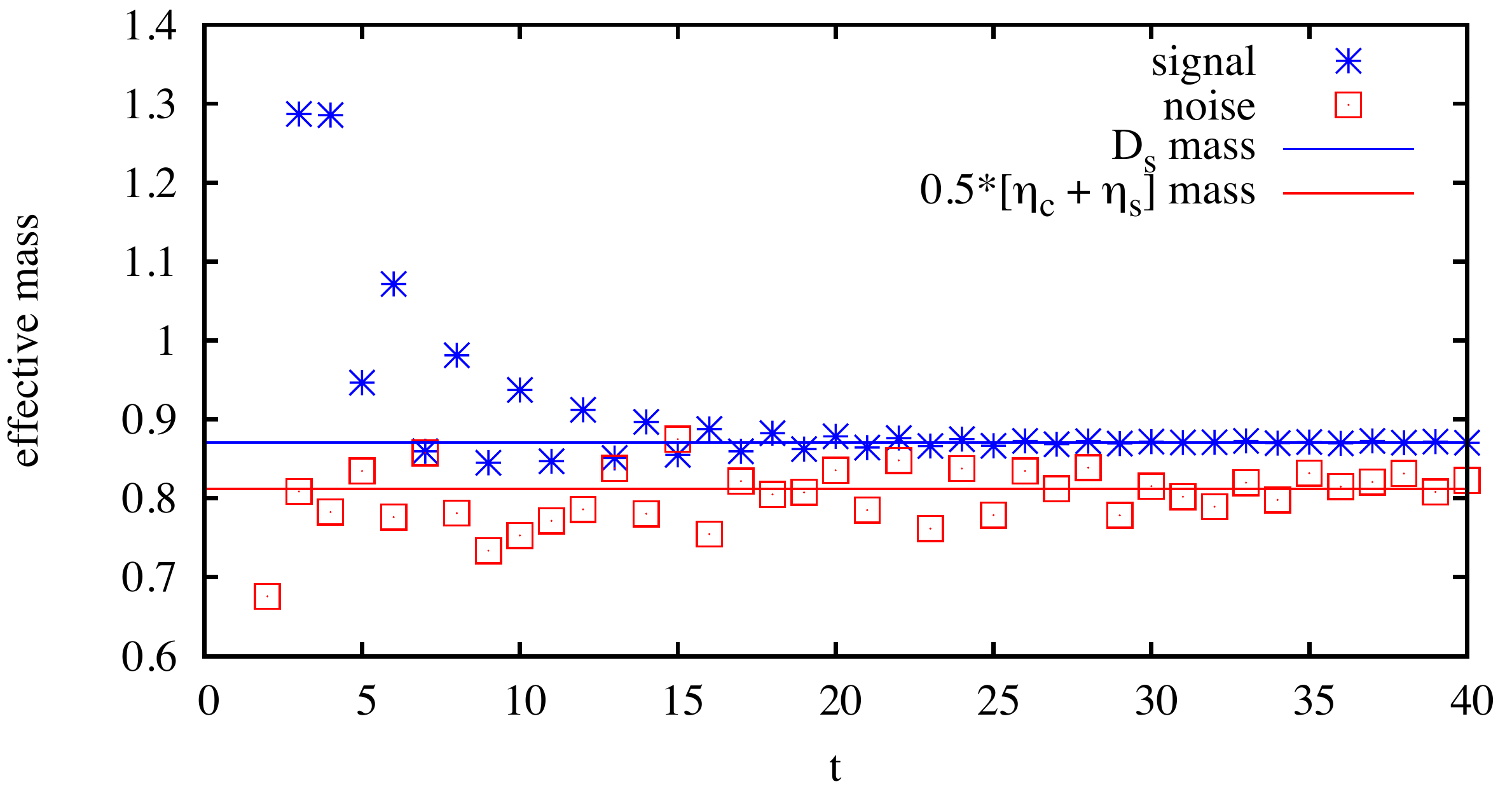}
\end{center}
\caption{Results for the effective mass of the $D_s$ 
correlator and the effective mass of the noise 
in the $D_s$ correlator plotted as a function of lattice 
time for one correlator on the fine lattices (set 8). 
The effective mass is obtained from the log of the ratio 
of the correlator (or its error) at successive times. 
At large times it becomes the mass of the lowest state 
in the correlator or its error. 
The lines compare the results to the expected mass i.e. 
the $D_s$ mass for the signal and $(m_{\eta_s}+m_{\eta_c})/2$ 
for the noise. 
}
\label{fig:massnoise}
\end{figure}

The $D_s$ meson correlators are made from the same 
$c$ and $s$ quark propagators that are used for 
the $\eta_c$ and $\eta_s$ above. We must use 
equation~\ref{eq:fit2} to fit the $D_s$ correlators,  
however, because they do have additional 
oscillating terms in them. 
Table~\ref{tab:ds} lists results for the masses, $M_0$ 
and the decay constant derived from $a_0$ for each 
combination of $c$ and $s$ masses that we have used 
on each ensemble. The statistical errors coming from 
the fit are significantly larger for the $D_s$ than 
for the $\eta_c$. This is because the noise in heavy-light 
correlators has a lower mass associated with it than 
the signal. The mass in the squared correlator which gives
the noise is given by one half of the sum of the $\eta_c$ and
$\eta_s$ masses, which is smaller than the signal $D_s$ 
mass. This means that the signal to noise 
ratio degrades at large times for the $D_s$ correlator 
and the statistical error increases. This is illustrated 
in Figure~\ref{fig:massnoise} in which we explicitly plot 
and compare the `effective mass' extracted from the $D_s$ correlator and 
from its statistical error. 
This issue becomes a problem for $B$ meson correlators~\cite{ericlat08}. 
It is not a big problem for the $D_s$, however, and the 
statistical errors that we obtain in Table~\ref{tab:ds} 
are very small. 

\begin{table}
\caption{\label{tab:ds}
Results for the mass and decay constant of the $D_s$ meson 
in units of the lattice spacing for 
a range of charm and strange quark masses on each MILC 
ensemble. }
\begin{ruledtabular}
\begin{tabular}{lllll}
Set & $am_c$ & $am_s$ & $am_{D_s}$ & $af_{D_s}$  \\
\hline
1 &  0.81 & 0.061 & 1.4665(8) & 0.1970(10) \\
 &  0.825 & 0.066 & 1.4869(7) & 0.1994(10) \\
 &  0.825 & 0.080 & 1.5019(6) & 0.2042(8) \\
 &  0.85 & 0.066 & 1.5117(8) & 0.2004(10) \\
 &  0.85 & 0.080 & 1.5266(6) & 0.2053(9) \\
2 &  0.825 & 0.066 & 1.4869(11) & 0.1997(20) \\
\hline
3 & 0.622 & 0.0489 & 1.1890(7) & 0.1538(9) \\ 
 &  0.65 & 0.0537 & 1.2247(5) & 0.1561(9)  \\
4 &  0.63 & 0.0492 & 1.2007(5) & 0.1559(7) \\
 &  0.66 & 0.0546 & 1.2391(5) & 0.1586(6) \\
 &  0.66 & 0.06 & 1.2452(5) & 0.1604(6) \\
 &  0.66 & 0.063 & 1.2486(4) & 0.1614(6) \\
 &  0.72 & 0.0546 & 1.3027(6) & 0.1602(7) \\
 &  0.72 & 0.06 & 1.3086(5) & 0.1620(7) \\
 &  0.72 & 0.063 & 1.3120(5) & 0.1631(6) \\
 &  0.753 & 0.0546 & 1.3369(6) & 0.1610(7) \\
 &  0.753 & 0.06 & 1.3429(5) & 0.1629(7) \\
 &  0.753 & 0.063 & 1.3462(5) & 0.1639(7) \\
5 &  0.63 & 0.0492 & 1.2013(5) & 0.1561(8) \\
6 &  0.625 & 0.0491 & 1.1916(7) & 0.1553(10) \\
7 &  0.619 & 0.0487 & 1.1867(10) & 0.1548(17) \\
\hline
8 &  0.413 & 0.0337 & 0.84721(23) & 0.10836(24)  \\
 &  0.43 & 0.0358 & 0.86982(23) & 0.10943(24)  \\
 &  0.43 & 0.0366 & 0.87079(22) & 0.10970(24)  \\
 &  0.43 & 0.0382 & 0.87274(21) & 0.11028(24)  \\
 &  0.44 & 0.0358 & 0.88152(23) & 0.10959(27)  \\
 &  0.44 & 0.0366 & 0.88249(23) & 0.10986(27)  \\
 &  0.44 & 0.0382 & 0.88443(22) & 0.11044(24)  \\
 &  0.45 & 0.0358 & 0.89317(24) & 0.10974(27)  \\
 &  0.45 & 0.0366 & 0.89414(23) & 0.11001(27)  \\
 &  0.45 & 0.0382 & 0.89607(23) & 0.11059(27)  \\
9 & 0.412 & 0.0336 &  0.84352(26) & 0.10779(31) \\
 &  0.427 & 0.03635 & 0.86443(40) & 0.1086(5) \\
\hline 
10 &  0.273 & 0.0228 & 0.59350(24) & 0.07500(27)  \\
\hline
11 & 0.193 & 0.0161 & 0.43942(33) & 0.05533(39) \\
 &  0.195 & 0.0165 & 0.44270(28) & 0.05550(34) \\
\end{tabular}
\end{ruledtabular}
\end{table}

To determine the physical mass of the $D_s$ meson 
as accurately as possible we want to minimise errors 
coming from the conversion from lattice units to 
physical units i.e. from the lattice spacing. 
The error on the physical value of $r_1$ is 0.7\%. 
Applied directly to the $D_s$ mass this would amount 
to a sizeable 14 MeV error. This can be avoided however, 
by calculating instead the mass difference 
$m_{D_s}-m_{\eta_c}/2$. Because this is much 
smaller (480 MeV) it will have a much reduced
absolute error from the lattice spacing~\cite{fds}. In addition, it is 
much less sensitive to any errors from 
mistuning of the $c$ quark mass because the 
leading contribution of $m_c$ effectively cancels 
in this difference. Indeed this difference can be thought of as
the difference in binding energy between a charmonium 
meson and a charm-light meson, and is therefore an 
important physical quantity. The fact that it 
can be calculated accurately in lattice QCD 
and compared to experiment is a stringent test 
of QCD itself.

\begin{figure}
\begin{center}
\includegraphics[width=80mm]{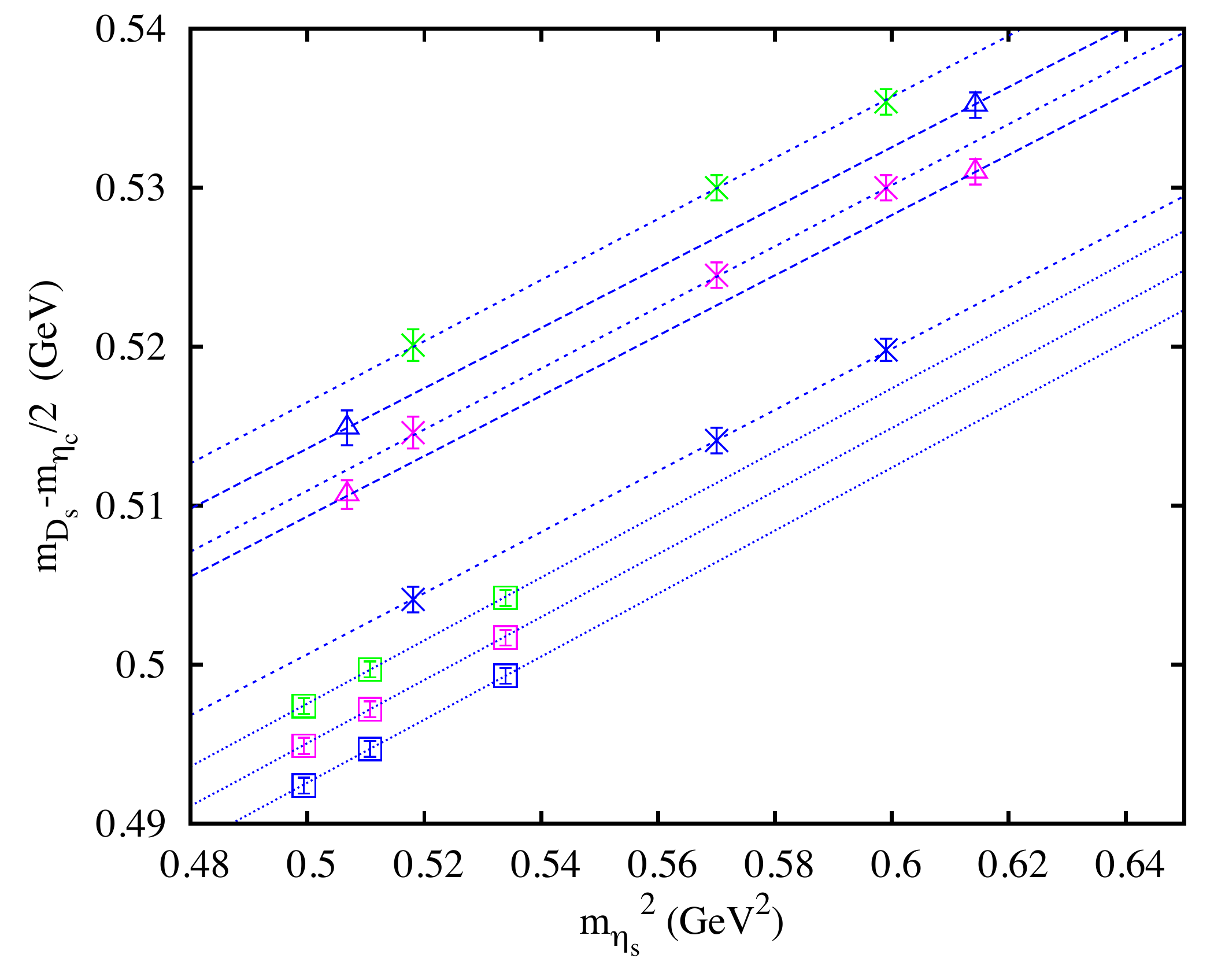}
\end{center}
\caption{Results for the mass of the $D_s$ meson 
(specifically the difference between that mass and one 
half of the $\eta_c$ mass) as a function of the square 
of the $\eta_s$ meson mass, acting as a proxy for the 
strange quark mass. 
Results are for a range of different quark masses around 
the masses of the $c$ and $s$ quark masses on very coarse 
set 1 (triangles), coarse set 4 (crosses) and fine set 7 (squares).
The lines are fits to the results for each ensemble allowing 
linear terms in $m_{\eta_s}^2$ and $m_{\eta_c}$. Here the lines 
join points for a fixed $c$ quark mass. See Figure~\ref{fig:mdstunec}
for the equivalent as a function of $m_{\eta_c}$. 
}
\label{fig:mdstunes}
\end{figure}

\begin{figure}
\begin{center}
\includegraphics[width=80mm]{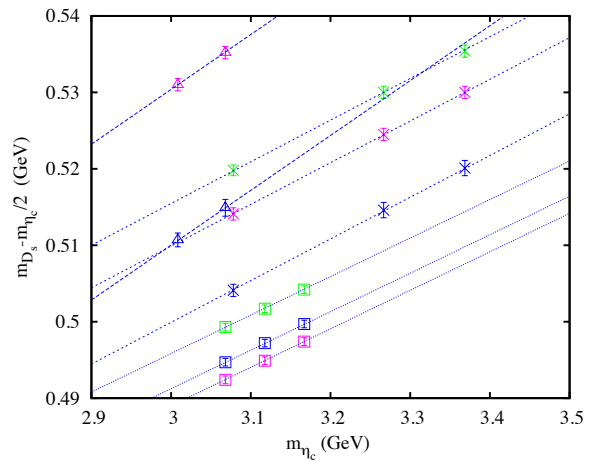}
\end{center}
\caption{Results for the mass of the $D_s$ meson 
(specifically the difference between that mass and one 
half of the $\eta_c$ mass) as a function of the 
$\eta_c$ meson mass, acting as a proxy for the 
charm quark mass. 
Results are for a range of different quark masses around 
the masses of the $c$ and $s$ quark masses on very coarse 
set 1 (triangles), coarse set 4 (crosses) and fine set 7 (squares).
The lines are fits to the results for each ensemble allowing 
linear terms in $m_{\eta_s}^2$ and $m_{\eta_c}$. Here the lines 
join points for a fixed $s$ quark mass. See Figure~\ref{fig:mdstunes}
for the equivalent as a function of $m_{\eta_s}^2$. 
}
\label{fig:mdstunec}
\end{figure}

The first stage in the analysis of the $D_s$ meson mass
is to determine the difference $m_{D_s} - m_{\eta_c}/2$ for 
tuned $c$ and $s$ quark masses on each ensemble. As discussed 
above, we have results very close to the tuned point on 
almost every one of the 11 ensembles. However, it is important 
to make sure that all of our results are tuned to the same point
before extrapolation and so we first test the dependence 
of $m_{D_s} - m_{\eta_c}/2$ as a function of $m_s$ and $m_c$. 
Figures~\ref{fig:mdstunes} and~\ref{fig:mdstunec} show 
results as a function of $m_{\eta_s}^2$ and $m_{\eta_c}$ 
for sets 1, 4 and 7 where we have multiple data points 
with different combinations of $m_c$ and $m_s$ and so 
can unravel the separate dependences. 
The dependence is plotted against meson mass rather than 
directly against the quark mass since the tuning condition is 
set by the $\eta_c$ or $\eta_s$ meson mass, so this is a 
more direct (and more physical) way to study any mistuning effects.  
Note that the mass values of the $\eta_c$ and $\eta_s$ 
are above their physical values for the cases given in 
figures~\ref{fig:mdstunes} and~\ref{fig:mdstunec}. Since 
we are only studying small mistuning effects for the 
values of the masses that we have closer to the physical 
points, this will give a sufficiently accurate picture of 
these effects. 

In figure~\ref{fig:mdstunes} we see that the dependence of 
$m_{D_s} - m_{\eta_c}/2$ on 
$m_{\eta_s}^2$ is linear as we expect, since this corresponds 
to a linear dependence on $m_s$. The slope is clearly physical 
i.e. independent of the lattice spacing (whereas the slope against 
$m_s$ would not be, because of the running of $m_s$ itself, 
discussed earlier). The value of the slope is 0.20(1) and this 
can be compared to an `experimental' slope, albeit over a 
much larger mass range, of 0.22 obtained 
by comparing results for the masses of the $D$ and the $D_s$~\cite{pdg}.  
Figure~\ref{fig:mdstunec} also shows linear dependence on 
$m_c$, expressed physically as linear dependence on $m_{\eta_c}$. 
The slope does differ on the very coarse lattices from the others
so showing some lattice spacing dependence in this case. 
The slope is also very small $\sim$ 0.05 because, as discussed 
above, the leading dependence on $m_c$ cancels between $m_{D_s}$ 
and $m_{\eta_c}/2$. The slope is again similar to the 
`experimental' value of 0.03 obtained over a much larger 
mass range from comparing $B_s$ and $D_s$ mesons~\cite{pdg}. 

Results from Figures~\ref{fig:mdstunes} and~\ref{fig:mdstunec} 
can be used to adjust the values of $m_{D_s}-m_{\eta_c}/2$ 
on each ensemble to the tuned point, $m_{\eta_c}$ = 2.985 GeV 
and $m_{\eta_s}$ = 0.6858 GeV. An error of 50\% of any shift 
is added in quadrature to the statistical error. The shifts 
from mistuning are less than the statistical error on all 
ensembles except sets 2 (very coarse) and 10 (superfine). On 
set 10 the shift is by 1.5 times the statistical error and 
on set 2 by 4 times the statistical error. 
Table~\ref{tab:tune} gives the tuned value of $m_{D_s}-m_{\eta_c}/2$ 
in GeV on each ensemble along with two errors. 
The first is the statistical/tuning error and the second 
is that from the error in $r_1/a$
on that ensemble. This error is a factor of 3 smaller than 
its naive value because of a cancellation of lattice spacing 
errors inside the mass difference. Any change in $r_1/a$ means
a change to $m_{\eta_c}$ and $m_{\eta_s}$ as well as a change 
in $m_{D_s}-m_{\eta_c}/2$. The results then need to be retuned 
to the physical $c$ and $s$ masses and this largely cancels 
the change resulting from the change in $r_1/a$. 
The error from $r_1/a$ uncertainty is much smaller than the 
statistical error then in every case. The statistical errors, 
which dominate, are at the level of 1 MeV.  

\begin{table}
\caption{\label{tab:tune}
Values for the mass and decay constant of the $D_s$ meson 
and for the decay constant of the $\eta_c$ after 
tuning to the physical $c$ and $s$ masses (i.e. 
the physical $\eta_c$ and $\eta_s$ meson masses) 
on each ensemble. Results are in GeV with two 
errors, the first from statistics and tuning and the 
second from the uncertainty in $r_1/a$ on that ensemble. 
}
\begin{ruledtabular}
\begin{tabular}{llll}
Set & $m_{D_s}-\frac{m_{\eta_c}}{2}$ (GeV)  & $f_{D_s}$ (GeV) & $f_{\eta_c}$ (GeV)  \\
\hline
1 &  0.5021(12)(4) & 0.2674(14)(3) & 0.4753(9)(2)  \\
2 &  0.5020(32)(4) & 0.2671(28)(3) & 0.4756(6)(2) \\
\hline
3 & 0.4889(12)(3) & 0.2564(15)(2) & 0.4284(3)(1) \\ 
4 &  0.4897(9)(3) & 0.2573(12)(2) & 0.4291(4)(1) \\
5 &  0.4906(9)(3) & 0.2576(13)(2) & 0.4292(4)(1) \\
6 &  0.4909(12)(3) & 0.2586(17)(2) & 0.4255(5)(1) \\
7 &  0.4911(17)(3) & 0.2592(28)(2) & 0.4286(6)(1) \\
\hline
8 &  0.4823(6)(2) & 0.2525(6)(2) & 0.4012(3)(2)  \\
9 & 0.4817(6)(2) & 0.2520(7)(2) &  0.3998(3)(2) \\
\hline 
10 &  0.4784(10)(2) & 0.2499(9)(3) & 0.3945(10)(3)  \\
\hline
11 & 0.4766(13)(4) & 0.2481(17)(5) & 0.3953(13)(6) \\
\end{tabular}
\end{ruledtabular}
\end{table}

We can then extrapolate the tuned values on each ensemble 
in the lattice spacing and the sea quark masses to the physical 
point where the lattice spacing is zero and the sea quark 
masses take their real world values. 
It is clear from Table~\ref{tab:tune} looking at the coarse 
and fine ensembles that $m_{D_s}-m_{\eta_c}/2$ 
has no significant dependence on the sea quark masses at 
the level of our 1 MeV statistical errors. 
The picture is obscured on the very coarse lattices 
by the larger error on set 2 from mistuning. In fact 
if we compare sets 1 and 2 at the $\eta_c$ and $\eta_s$ 
masses corresponding to those available on set 2 
(i.e. at somewhat heavier masses than the correctly 
tuned point) 
then we find again that sets 1 and 2 agree on the 
value of $m_{D_s}-m_{\eta_c}$ but now within an 
error of 1.5 MeV rather than the 3 MeV in Table~\ref{tab:tune}. 

We expect $m_{D_s}-m_{\eta_c}/2$ to be very insensitive 
to the sea quark masses based on chiral perturbation theory. 
This couples a nonrelativistic lagrangian for $D$ and $D_s$ meson fields 
to the pion octet and gives an expansion in powers of $\pi$, $K$
and $\eta_8$ masses for the mass and decay constant of the 
appropriate $D$ meson. The $D_s$ has valence $c$ and $s$ masses 
which have been tuned to the appropriate values so the only 
dependence we are interested in here is the dependence on 
sea $s$ and $u/d$ quark masses which enter through the masses 
of mesons made either purely of sea quarks or of mixed sea and 
valence quarks. 
The leading tree-level dependence on sea quark masses 
is a term $C(2m_{l,sea}+m_{s,sea})$. Loops couple the $D_s$ meson to a 
virtual $DK$ or $D_s \eta_8$ pair.  This generates logarithmic 
terms but with, in this case, a very benign dependence on sea 
quark masses since none of the associated meson masses vanish 
in the chiral limit. These terms can then simply be viewed 
as additional polynomial terms in $m_{l,sea}$ and $m_{s,sea}$.  
A more detailed chiral analysis is not useful here because 
the sea quark mass dependence of our results is clearly so 
small as to have no useful information in it. We simply need 
to make sure that we allow a sufficient error on the extrapolated 
value at the physical point to allow for any 
sea quark mass dependence that might be there. 
For this purpose a simple polynomial expansion in $m_{l,sea}$ and 
$m_{s,sea}$ suffices. We take as expansion coefficients 
$\delta x_{l}$ and $\delta x_{s}$ where $\delta x_{q} = (m_{q,sea} - m_{q, sea, phys})/m_{s, sea, phys}$. 
$m_{s,sea,phys}$ is the sea (asqtad) strange quark mass 
at the physical point. We take this value from results quoted 
by the MILC collaboration~\cite{milcreview} for very coarse to superfine
 and use the analysis of the ratio 
of HISQ to asqtad mases from Appendix~\ref{app:sea} to 
give the value of $m_{s,sea,phys}$ on the ultrafine 
lattices. We take $m_{l, sea, phys} = m_{s, sea, phys}/27.2$ using
the ratio for $m_l/m_s$ determined by the MILC collaboration~\cite{milcreview}. 
Table~\ref{tab:tune} shows that $m_{D_s} - m_{\eta_c}/2$ does 
have significant dependence on the lattice spacing, changing 
by 20 times the statistical error between very coarse and 
ultrafine lattices. 
This is also not surprising because the charm quark 
is relatively heavy and consequently the scale for discretisation 
errors here will be much higher than that for quantities involving 
only light quarks. This is why it is important to have a formalism, 
such as HISQ, with very well controlled discretisation errors 
and to have results at many values of the lattice spacing. 
Discretisation errors with the HISQ action can appear only as 
powers of $a^2$ - no odd powers of $a$ are allowed. The $a^2$ errors 
appearing at tree level have been removed and so the coefficient 
of $a^2$ terms is $\cal{O}$$(\alpha_s)$. The inclusion of the 
Naik term with coefficient calculated at tree-level means that 
all $(m_ca)^{2n}$ discretisation errors are removed at leading 
order in $v^2/c^2$ where $v^2$ is the velocity of the charm 
quark in the $D_s$ or $\eta_c$. Thus discretisation errors 
from the HISQ action are expected to be at the level 
of $(v^2/c^2)(m_ca)^{2n}$, except for the $a^2$ term which 
is further suppressed by $\alpha_s$. There are additional 
$\alpha_s a^2$ and tree-level $a^4$ and higher 
errors coming from the gluon action, however. These we would 
typically expect to have a scale of a few hundred MeV (i.e. 
$\Lambda_{QCD}$) associated with them rather than $m_c$, so 
their effects will be included if we allow for a scale of $m_c$. 
  
We therefore take the following fit form to extrapolate 
$\Delta = m_{D_s}-m_{\eta_c}/2$ to the physical point:
\begin{eqnarray}
\label{eq:fitform}
\Delta(a,\delta x_l, \delta x_s) &=& \Delta_{phys}[1+\sum_{j=1}^4 c_j(m_ca)^{2j} \\ \nonumber 
&+& 2b_l\delta x_l(1+c_b(m_ca)^2) \\ \nonumber
&+& 2b_s\delta x_s(1+c_s(m_ca)^2) \\ \nonumber
&+& 4b_{ll}(\delta x_l)^2 + 2b_{ls}\delta x_l\delta x_s + b_{ss}(\delta x_s)^2].
\end{eqnarray} 
We use a constrained fit~\cite{gplbayes} to this form 
which allows us to estimate 
the errors arising from different pieces of the fit. 
The prior value and width for $\Delta_{phys}$ we take as 0.5, with the 
very broad width of 0.2.  
Note that we give the discretisation errors a scale of $m_c$. The prior 
value and width that we take on the $c_n$ parameters is 0.0(2), estimating 
$v^2/c^2$ for the $c$ quark inside the $D_s$ to be 0.2. 
$c_1$, which multiplies the $a^2$ errors, is a factor of $\alpha_s$ 
smaller from the arguments above so we take the prior for $c_1$ to 
be 0.00(6). 
The $b$ parameters multiplying the linear sea quark mass dependence are taken 
to have prior values and widths of 0.00(7). The size of the prior width 
here is set by the fact that the dependence of $\Delta$ on the 
valence light quark mass inside the $D_s$ is known from a comparison 
of $D$ and $D_s$. This would give a slope with valence mass, in units 
of the strange mass, of 0.2. 
Sea quark mass effects 
are a factor of at least 3 smaller than valence mass effects in 
gold-plated quantities, so we take a prior width of 0.07. 
By the same 
reasoning we allow the $b$ parameters
multiplying the quadratic dependence to be  as large as
 $(0.2)^2/3$, i.e. we take the prior on these parameters to be 0.000(13).

The extrapolated result at the physical point, $\Delta_{phys}$ 
from the fit above is 0.4753(22) MeV with a $\chi^2/{\rm dof}$ of 0.2 for 
11 degrees of freedom. We fit all of the data including the two 
volumes for the coarse lattices, sets 4 and 5. Missing out set 
5 makes no appreciable difference to the result. 
Modifications to the fit 
form above also do not change this number significantly. Here we itemize 
the effect of some of them: 
\begin{itemize}
\item changing the prior on all $c_i$ (including $c_1$) 
to 0.0(5) changes $\Delta_{phys}$  
0.4$\sigma$ and increases the error by 20%. 
\item adding two extra powers of $a^2$ into the sum on $j$ in 
equation~\ref{eq:fitform} (i.e. using 6 terms instead of 4) 
does not change $\Delta_{phys}$ or
the error at all. The same is true for subtracting two 
powers of $a^2$ (i.e. using 2 terms instead of 4).  
\item adding extra discretisation errors into the sea-quark mass 
dependence (i.e. a term proportional to $(m_ca)^4$ in each of 
the terms linear in $\delta x_l$ and $\delta x_s$ and a term 
proportional to $(m_ca)^2$ in each of the quadratic terms) makes no 
difference at all. 
\item  missing out the sea quark mass dependence altogether makes 
no difference to $\Delta_{phys}$ but increases the 
$\chi^2$ value to 0.33. 
\item Changing all the $\delta x$ values 
by 10\% in either direction makes no appreciable difference, nor does 
changing them within their error bars on, for example, the 
ultrafine or fine lattices. 
\item missing out the very coarse lattice results makes no difference; 
missing out the very coarse and the coarse shifts 
$\Delta_{phys}$ by 0.4$\sigma$ (1 MeV), and increases the error to 3 MeV 
as $\chi^2$ drops to 0.1.  
\item missing out the ultrafine result shifts $\Delta_{phys}$ by 0.2$\sigma$ 
(0.5 MeV) and increases the error to 3 MeV. 
\end{itemize}

\begin{figure}
\begin{center}
\includegraphics[width=80mm]{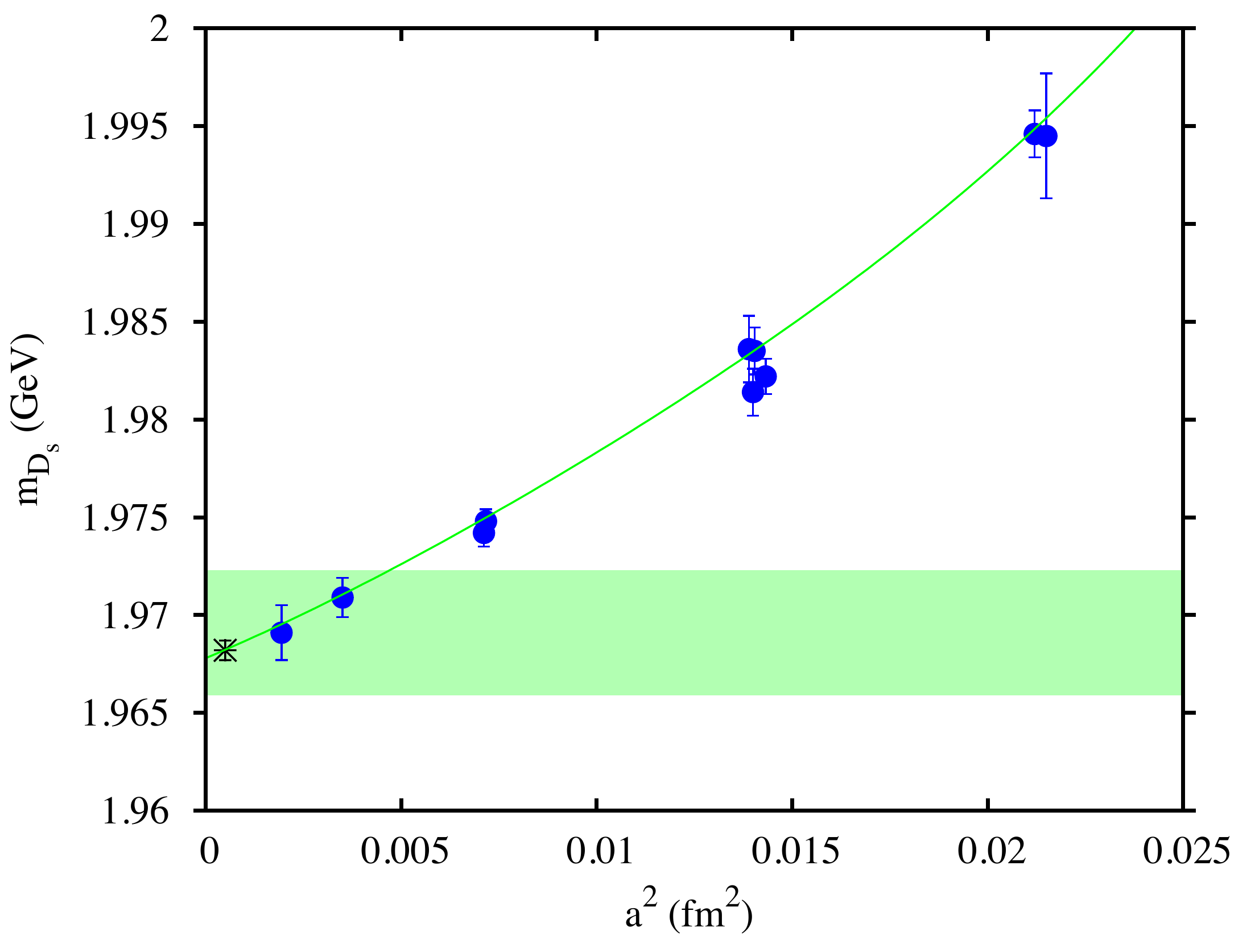}
\end{center}
\caption{Results for the mass of the $D_s$ meson 
tuned to the correct valence $c$ and $s$ mass on each 
ensemble from Table~\ref{tab:tune} as a function of the square of the lattice 
spacing. 
The line shows the result of the fit 
described by equation~\ref{eq:fitform}, taken at the physical values for 
the sea quark masses.  
The shaded band gives our final result adjusted for electromagnetic 
effects and with the full error as described in the text. 
The black burst gives the experimental result. 
}
\label{fig:mdsvsasq}
\end{figure}

Figure~\ref{fig:mdsvsasq} shows the results plotted against 
the square of the lattice spacing along with the fitted curve above, 
taken at the physical sea quark mass values (i.e. $\delta x_l = \delta x_s = 0$). 
The value plotted on the $y$-axis is $m_{D_s}$ itself, generated 
by adding $m_{\eta_c}/2$ = 1.4925 GeV to $\Delta$. 
The result at $a=0$ is then the value of the $D_s$ mass in a world 
without electromagnetism. To compare to experiment we need 
to estimate and add in the effect of the electromagnetic repulsion 
between the positively charged quark and antiquark inside the $D_s$. 
To do this we compare experimental masses for the $D^+$, $D^0$, $D_s$, 
$B^+$, $B^0$ and $B_s$ to a phenomenological formula allowing 
for electromagnetic effects proportional to the product of quark and 
antiquark electric charges inside the meson as well as the square 
of the electromagnetic charge on the light quark. This latter term 
is a self-energy effect, not needed for the heavy quarks because it
will cancel in all the differences taken (and therefore is absorbed 
into the heavy quark mass). In comparing charged and neutral 
mesons containing $u$ and $d$ quarks we must allow for the mass 
difference between $u$ and $d$ quarks. 
Then we can write~\cite{goity}: 
\begin{equation}
M(Q,q) = M_{sim}(Q,q) + A e_qe_Q + B e_q^2  + C (m_q - m_l)
\end{equation}
where $M_{sim}$ is the mass of the meson in the absence of 
electromagnetism and with $m_u = m_d$. 
If we take experimental results for the meson masses above 
along with $m_s/m_l = 27.2$ and $m_u/m_d = 0.42$ we obtain
$A \approx$ 4 MeV, $B \approx$ 3 MeV and $Cm_s \approx$ 100 MeV. 
The latter quantity differs by 10\% between $D$ and $B$ mesons, 
indicating $1/m_Q$ effects at this level that we ignore here. 
The resulting electromagnetic 
shift for the $D_s$ is then 1.3(7) MeV, where we take an error 
of 50\% on the shift, safely encompassing $1/m_Q$ effects and 
other limitations of this model. 
Adding 1.3 MeV to our fit result gives the shaded band in 
Figure~\ref{fig:mdsvsasq}, where we now include our full 
error of 3.2 MeV. The full error budget is discussed below.  

\begin{figure}
\begin{center}
\includegraphics[width=80mm]{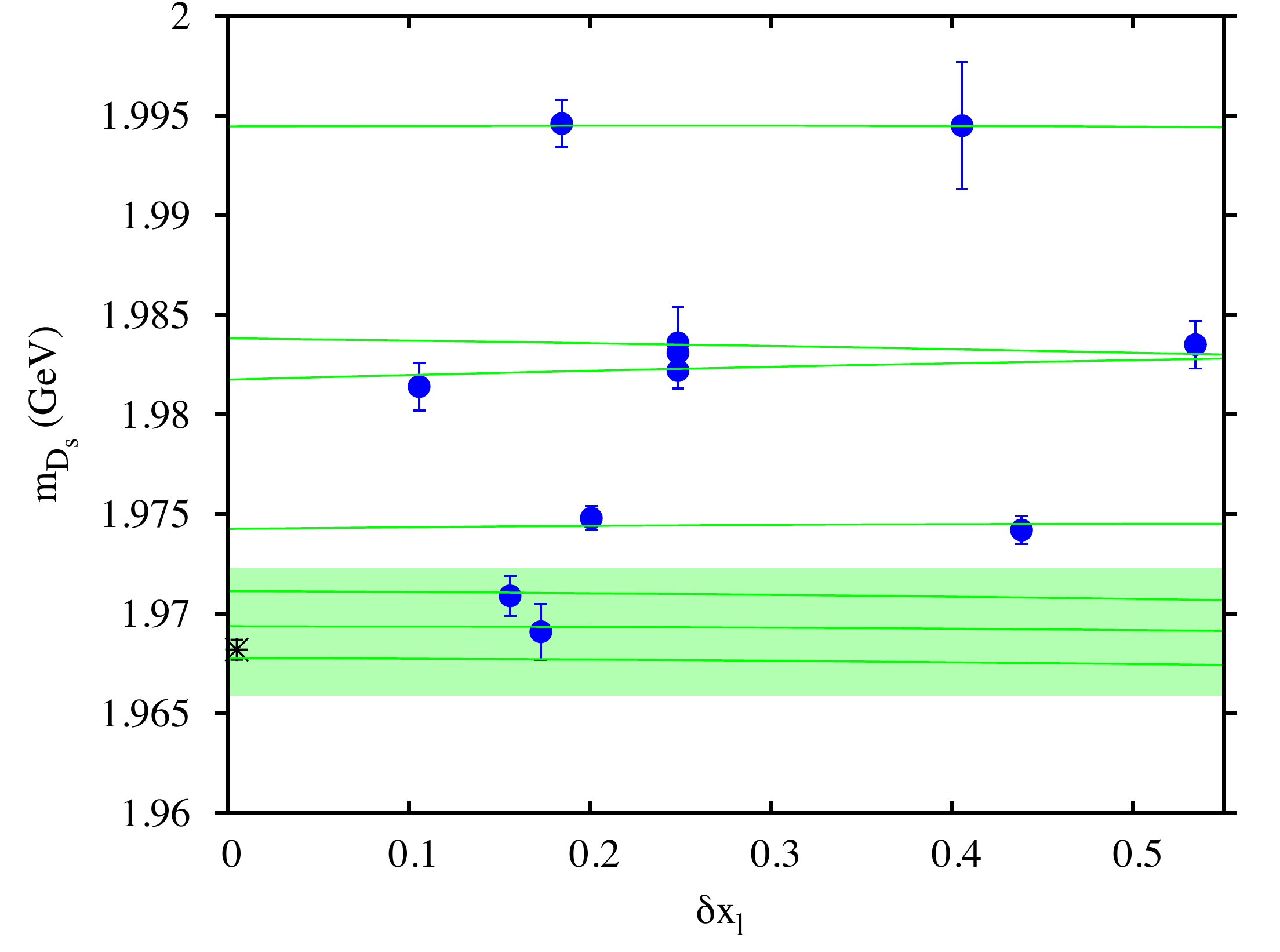}
\end{center}
\caption{Results for the mass of the $D_s$ meson 
tuned to the correct valence $c$ and $s$ mass on each 
ensemble from Table~\ref{tab:tune} as a function of the difference 
between the sea light quark mass and the physical value scaled 
by the physical strange quark mass (i.e. the parameter $\delta x_l$). 
The results are clearly separated by their lattice spacing value
with very coarse at the top and ultrafine at the bottom. 
The lines show the result of the fit 
described by equation~\ref{eq:fitform}, taken at the 
value of the sea strange quark mass ($\delta x_s$) and using the 
lattice spacing value corresponding to the ensemble with 
smallest $\delta x_l$ in that group. 
The results on the coarse lattices at $\delta x_l = 0.25$ 
include numbers at two different values of $\delta x_s$ as 
well as at two different volumes. This gives an idea of 
the spread in results from these effects. 
The lowest line is the fit curve in $\delta x_l$ at $a=0$
and $\delta x_s = 0$. 
The shaded band gives our final result adjusted for electromagnetic 
effects and with the full error as described in the text. 
The black burst gives the experimental result. 
}
\label{fig:mdsvsmsea}
\end{figure}

Figure~\ref{fig:mdsvsmsea} shows the sea quark mass 
dependence of our results plotted against $\delta x_l$.
The fitted curves are those from equation~\ref{eq:fitform}. 
For each group of ensembles we use the lattice 
spacing value from the ensemble with 
lightest sea quark mass to plot the fit curve. 
No significant dependence on $\delta x_l$ or $\delta x_s$ is evident.

\begin{table}
\caption{\label{tab:errors}
Full error budget for $m_{D_s}$, $f_{D_s}$ and 
$f_{\eta_c}$ given as a percentage of the final 
fitted value. Note that in the case of $f_{\eta_c}$ 
the top six errors are those to be considered for 
a lattice QCD calculation that matches this one. 
As discussed in the text, the bottom three errors are 
included for completeness. 
}
\begin{ruledtabular}
\begin{tabular}{llll}
Error & $m_{D_s}$ & $f_{D_s}$ & $f_{\eta_c}$ \\
\hline
statistical/valence tuning & 0.094\%  & 0.57\% & 0.45\%  \\
$r_1/a$ & 0.025\%  & 0.15\% & 0.16\% \\
$r_1$ & 0.051\% & 0.57\% & 0.27\% \\ 
$a^2$ extrapoln & 0.044\%  & 0.40\% & 0.24\% \\
$m_{q,sea}$ extrapoln & 0.048\% & 0.34\% & 0.09\% \\
finite volume & 0\% & 0.10\% & 0\% \\
$m_{\eta_s}$ & 0.056\%  & 0.13\% & -- \\
em effects in $D_s$ & 0.036\% & 0.10\%  &  -- \\
em and annihln in $m_{\eta_c}$ & 0.076\%  & 0.00\% & 0.05\% \\
em effects in $\eta_c$ & -- & -- & 0.40\% \\
missing $c$ in sea &  0.01\% & 0\% & 0.01\%  \\
\hline
{\bf Total} & 0.16\% & 1.0\% &  0.6\% (top 6)
\end{tabular}
\end{ruledtabular}
\end{table}

Table~\ref{tab:errors} shows the complete error budget for 
$m_{D_s}$ from our calculation.  The error of 2.2 MeV from 
our fit to $\Delta$ above includes the effect of statistical 
errors (including valence mass mistuning errors), $r_1/a$ errors
and errors arising from the extrapolation in sea quark masses 
and lattice spacing.  We can separate these errors  as
described in~\cite{hisq2} by working out how the final error 
changes when any of the inputs to the fit changes 
and dividing $\sigma^2$ into a sum of 
terms coming from each input: 
\begin{equation}
\sigma^2 = \sigma^2_a + \sigma^2_b + \ldots.
\end{equation}
Inputs to the fit include groups of priors associated 
with pieces of the fit function as well as statistical errors 
on the data points. 
Here we streamline the process by calculating explicitly the 
differential of $\chi^2$ with respect to the inputs and 
so determining $\sigma_a^2$, $\sigma_b^2$ etc. directly. 
The resulting breakdown of errors given in Table~\ref{tab:errors}
shows them to be dominated by statistical errors. 

Additional errors to be included in the error budget are 
errors that affect the final result in physical units but
do not affect the fit above. The first of these is the 
overall error in the physical value of $r_1$ of 0.7\%. This affects 
the tuning of all the valence masses but, as described earlier, 
the effect on $\Delta$ is reduced by a factor of 3 because 
of cancellation between scale shifts and tuning shifts. More precisely
we find a 1.0 MeV error on $m_{D_s}$ from the $r_1$ 
uncertainty. The effect on $\Delta$ of the 
uncertainty in the physical values of 
the $\eta_c$ mass and the $\eta_s$ mass used in tuning 
can be judged from 
Figures~\ref{fig:mdstunec} and~\ref{fig:mdstunes}. 
The error on the $\eta_c$ mass has negligible effect, again 
because most of the $\eta_c$ mass dependence cancels out 
in $\Delta$. The uncertainty in the $\eta_s$ mass is not 
negligible, however, but gives an uncertainty in 
$\Delta$, which we then transfer to $m_{D_s}$, of 1.1 MeV. 
The error on the physical value of the $\eta_c$ reappears 
when we reconstruct $m_{D_s}$ from $\Delta$ and 
$m_{\eta_c}$. It therefore gives a 1.5 MeV error to 
$m_{D_s}$ coming from electromagnetic and annihilation 
effects in the $\eta_c$ meson mass. The error from 
electromagnetic effects on the $D_s$ mass itself is 0.7 MeV 
as described earlier. 

The error from the finite volume of the lattices we estimate 
to be negligible from finite volume chiral perturbation theory. 
Our lattice results comparing two different volumes (sets 4 and 5) show 
no significant effect at the level of 0.4% statistical errors. 

Our lattice calculation includes $u$, $d$ and $s$ quarks in 
the sea but no $c$ quarks, although gluon field configurations 
are now being generated that do include them~\cite{newmilcdyn}. In the real world 
$c$ quarks do appear in the sea and we can estimate the 
effect of these perturbatively because the $c$ quark mass 
is relatively heavy, i.e. larger than typical momenta appearing 
inside the mesons we are discussing. The effect of a massive 
quark loop in the gluon propagator which gives rise to the 
heavy quark potential is simply to add a correction to the 
potential which is proportional to a delta function at the 
origin~\cite{itz}:  
\begin{equation}
V(r) = -\frac{C_f \alpha_s}{r} \rightarrow - C_f \alpha_s \left(\frac{1}{r} + 
\frac{\alpha_s}{10m_c^2} \delta^3(r)\right). 
\end{equation}
Although this additional term is a spin-independent interaction its 
effects in charmonium can be judged by comparison to that of the hyperfine 
potential. The hyperfine potential induces a mass splitting of 
$\approx 120$ MeV from a term which has the same $\delta$ function 
form as above but a coefficient 280 (= $80 \pi/(3\alpha_s)$ ) times as large.  
Thus we expect the shift of the $\eta_c$ (and $J/\psi)$ masses 
caused by the presence of $c$ quarks in the sea to 
be approximately 0.4 MeV. The $D_s$ meson has much smaller 
momenta typically inside it and so we expect a much smaller effect 
from $c$ quarks in the sea on 
the $D_s$ meson mass. If we set that effect to zero, so that 
conservatively there is no cancellation of this effect in 
the quantity $\Delta$, then we obtain 
an uncertainty in our final $D_s$ mass of 0.2 MeV, or 0.01\%. 

Our final result for $m_{D_s}$ is then 1.9691(32) GeV to be compared 
to an experimental result of 1.9685(3) GeV~\cite{pdg}. 

\subsection{$f_{D_s}$}
\label{sec:fds}

The decay constant of the $D_s$ meson is the main result 
from this paper. Having discussed in detail the tests that 
can be successfully done of the $D_s$ mass, we now discuss the analysis 
of the decay constant. 

\begin{figure}
\begin{center}
\includegraphics[width=80mm]{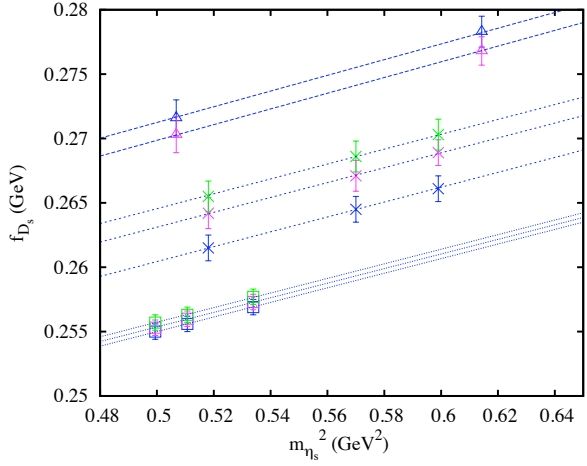}
\end{center}
\caption{Results for the decay constant of the the $D_s$ meson 
 as a function of the square 
of the $\eta_s$ meson mass, acting as a proxy for the 
strange quark mass. 
Results are for a range of different quark masses around 
the masses of the $c$ and $s$ quark masses on very coarse 
set 1 (triangles), coarse set 4 (crosses) and fine set 7 (squares).
The lines are fits to the results for each ensemble allowing 
linear terms in $m_{\eta_s}^2$ and $m_{\eta_c}$. Here the lines 
join points for a fixed $c$ quark mass. See Figure~\ref{fig:fdstunec}
for the equivalent as a function of $m_{\eta_c}$. 
}
\label{fig:fdstunes}
\end{figure}

\begin{figure}
\begin{center}
\includegraphics[width=80mm]{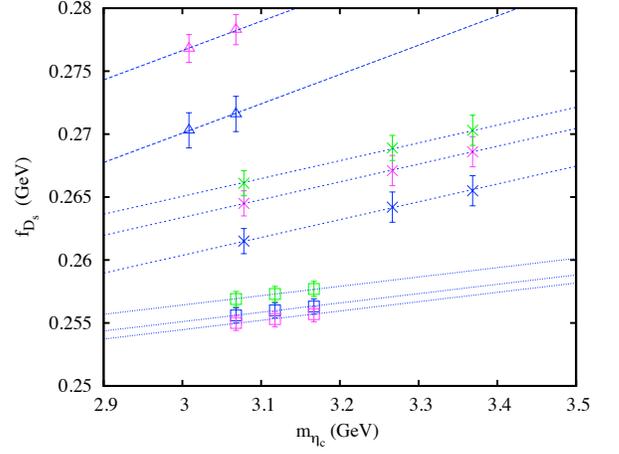}
\end{center}
\caption{Results for the decay constant of the the $D_s$ meson 
as a function of the 
$\eta_c$ meson mass, acting as a proxy for the 
charm quark mass. 
Results are for a range of different quark masses around 
the masses of the $c$ and $s$ quark masses on very coarse 
set 1 (triangles), coarse set 4 (crosses) and fine set 7 (squares).
The lines are fits to the results for each ensemble allowing 
linear terms in $m_{\eta_s}^2$ and $m_{\eta_c}$. Here the lines 
join points for a fixed $s$ quark mass. See Figure~\ref{fig:fdstunes}
for the equivalent as a function of $m_{\eta_s}^2$. 
}
\label{fig:fdstunec}
\end{figure}

Table~\ref{tab:ds} gives the raw results for the decay constant 
on the 11 different ensembles we have studied. As for $m_{D_s}$ 
it is important to be able to understand the dependence of 
$f_{D_s}$ on the valence $c$ and $s$ masses and to tune the 
result on each ensemble to the physical values for these masses. 
As described above, this corresponds to tuning them to physical 
values of the $\eta_c$ and $\eta_s$ meson masses.  Figures~\ref{fig:fdstunes}
and~\ref{fig:fdstunec} show the dependence of $f_{D_s}$ on these
meson masses on very coarse, coarse and fine lattices. Again 
we are using results somewhat above the physical values for 
the $\eta_s$ and $\eta_c$ masses
to extract the dependence which will then allow us to tune 
accurately our results that are much closer to the physical values.  
As expected, the dependence on $m_{\eta_s}^2 \equiv m_s$ 
is linear and the slope does not change with lattice spacing.   
The value of the slope, $0.06 {\rm GeV}^{-1}$ can be compared 
to the change in $f_{D_q}$ expected from $q=s$ to $q=l$~\cite{fds}. 
This corresponds to a somewhat larger slope of $0.09 {\rm GeV}^{-1}$ 
but is over a much larger range where nonlinear effects 
may appear.  The slope of $f_{D_s}$ against $m_{\eta_c}$ falls
from very coarse to fine lattices.  This has interesting 
implications for the behaviour of the heavy-strange meson 
decay constant as a function of heavy quark mass. It is clear 
from the study of the $\eta_s$ and $D_s$ mesons that the 
decay constant increases as the `heavy' quark mass is increased 
from $m_s$ to $m_c$. However, above $m_c$ the behaviour is less 
clear because lattice QCD calculations have so far not been 
accurate enough to distinguish clearly what is happening to 
within 5-10\% errors. 
There are known to be
large corrections to the $1/\sqrt{m_Q}$ behaviour expected from 
HQET because $f_{D_s}$ and $f_{B_S}$ are not very different~\cite{ourfbs}. 
This is consistent with a slope against heavy quark mass for 
$f_{D_s}$ that tends to zero.  It is clear that understanding 
this dependence also requires good control of discretisation errors. 

Again we use the dependence shown in these plots 
to make small tuning shifts 
to the values of $f_{D_s}$ on each ensemble so that they 
correspond to the correct result for $m_{\eta_c}$ = 2.985 GeV 
and $m_{\eta_s}$ = 0.6858 GeV. Table~\ref{tab:tune} gives 
the tuned values on each ensemble. Because the statistical errors 
are about twice as large for $f_{D_s}$ as for $\Delta$ and 
the dependence on $m_{\eta_c}$ and $m_{\eta_s}$ is smaller, 
the tuning shifts, and the errors from them, are very much 
less than the statistical errors on all ensembles. Even on set 
2 the shift from mistuning is only 1$\sigma$. The 
error in $f_{D_s}$ from the uncertainty in $r_1/a$ is only slightly 
reduced over its naive value from cancellations. It is also much smaller 
than the statistical error everywhere. It is given as the 
second error in Table~\ref{tab:tune}.  

Again it is clear from Table~\ref{tab:tune} that the sea quark 
mass dependence of the results is smaller than our 1-2 MeV 
statistical errors, but the lattice spacing dependence is not. 
We therefore fit the sea quark mass dependence with a relatively 
simple form that allows an error for what little dependence 
there is to be included in the final 
extrapolated value at the physical point. 
For the lattice spacing dependence we include 
relatively high order terms to make sure that a sufficiently large 
error is included in the final extrapolated value for this 
dependence. The fit form is the same as that used for $\Delta$: 
\begin{eqnarray}
\label{eq:fitformf}
f_{D_s}(a,\delta x_l, \delta x_s) &=& f_{D_s,phys}[1+\sum_{j=1}^4 c_j(m_ca)^{2j} \\ \nonumber 
&+& 2b_l\delta x_l(1+c_b(m_ca)^2) \\ \nonumber
&+& 2b_s\delta x_s(1+c_s(m_ca)^2) \\ \nonumber
&+& 4b_{ll}(\delta x_l)^2 + 2b_{ls}\delta x_l\delta x_s + b_{ss}(\delta x_s)^2].
\end{eqnarray} 
We take the same prior values and widths as before except that 
for $f_{D_s,phys}$ we take to be 0.25(10). 

The extrapolated result at the physical point, $f_{D_s,phys}$ is 
0.2480(19) GeV with a $\chi^2/{\rm dof}$ of 0.2 for 11 degrees 
of freedom. The fit is robust to changes in the fitting function:
\begin{itemize}
\item changing the prior on all the $c_i$ (including $c_1$) 
to 0.0(5) changes $f_{D_s,phys}$ by 
0.8$\sigma$ and increases the error by 30\%. 
\item adding or subtracting two powers of $a^2$ into the sum on $j$ in 
equation~\ref{eq:fitformf} does not change $f_{D_s,phys}$ or its error.  
\item adding an extra power of discretisation errors into both the 
linear and quadratic sea-quark mass 
dependent terms makes no difference. 
\item  missing out the sea quark mass dependence altogether 
changes $f_{D_s,phys}$ by 0.2$\sigma$ but increases the 
$\chi^2$ value to 0.3. 
\item Changing all the $\delta x$ values 
by 10\% in either direction makes no appreciable difference, nor does 
changing them within their error bars on, for example, the 
ultrafine or fine lattices. 
\item missing out the very coarse lattice results does not 
change $f_{D_s, phys}$; 
missing out the very coarse and the coarse shifts 
$f_{D_s,phys}$ by 0.3$\sigma$ (1 MeV). 
\item missing out the ultrafine result shifts $f_{D_s, phys}$ by 0.4$\sigma$ 
(1 MeV).  
\end{itemize} 

\begin{figure}
\begin{center}
\includegraphics[width=80mm]{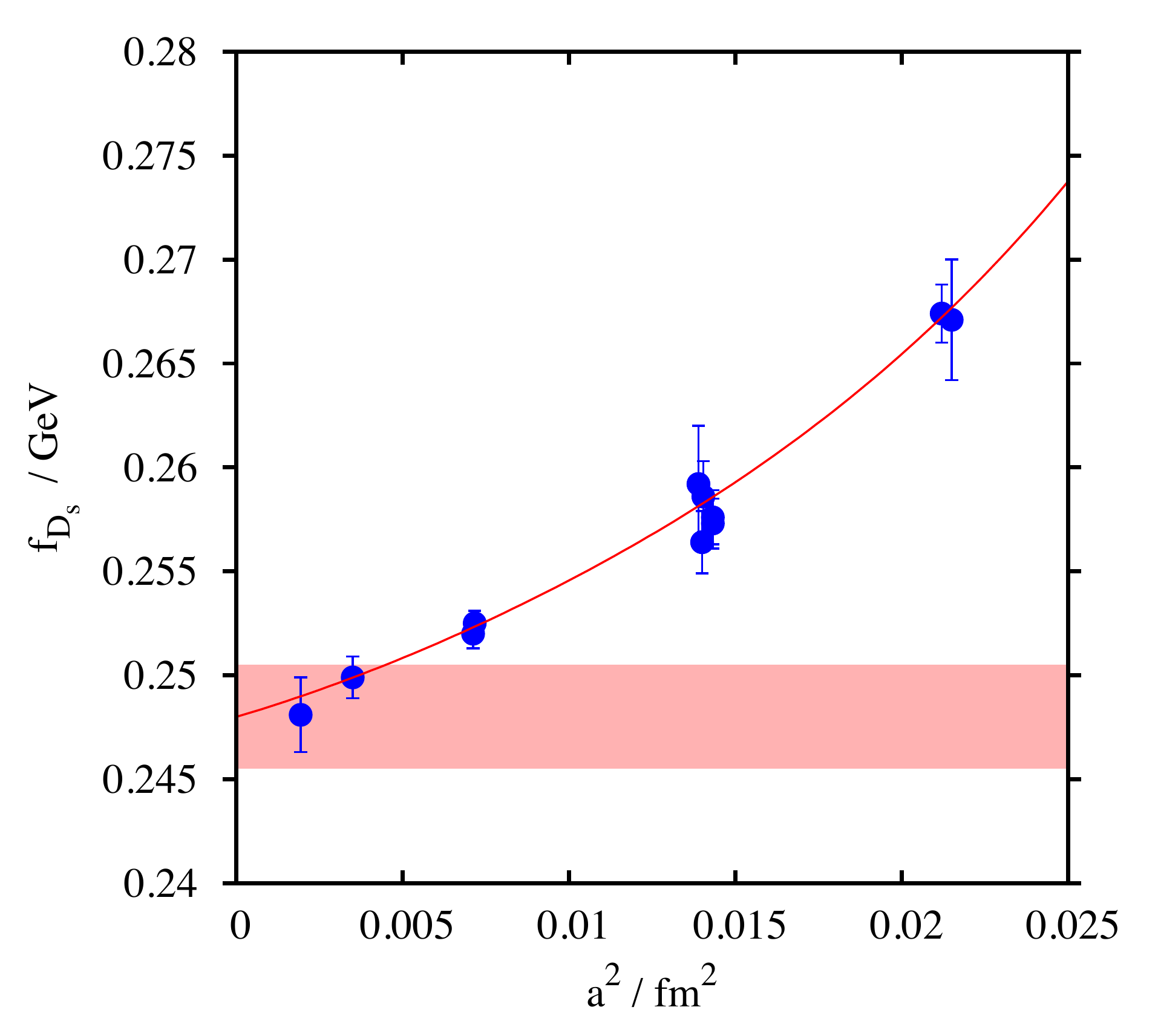}
\end{center}
\caption{Results for the $D_s$ decay constant
tuned to the correct $c$ and $s$ mass on each 
ensemble as a function of the square of the lattice 
spacing. The line shows the result of the fit 
at the physical value for the sea quark masses, 
as described in the text.  
The shaded band gives our final result with the 
full error bar as described in the text. 
}
\label{fig:fdsvsasq}
\end{figure}

\begin{figure}
\begin{center}
\includegraphics[width=80mm]{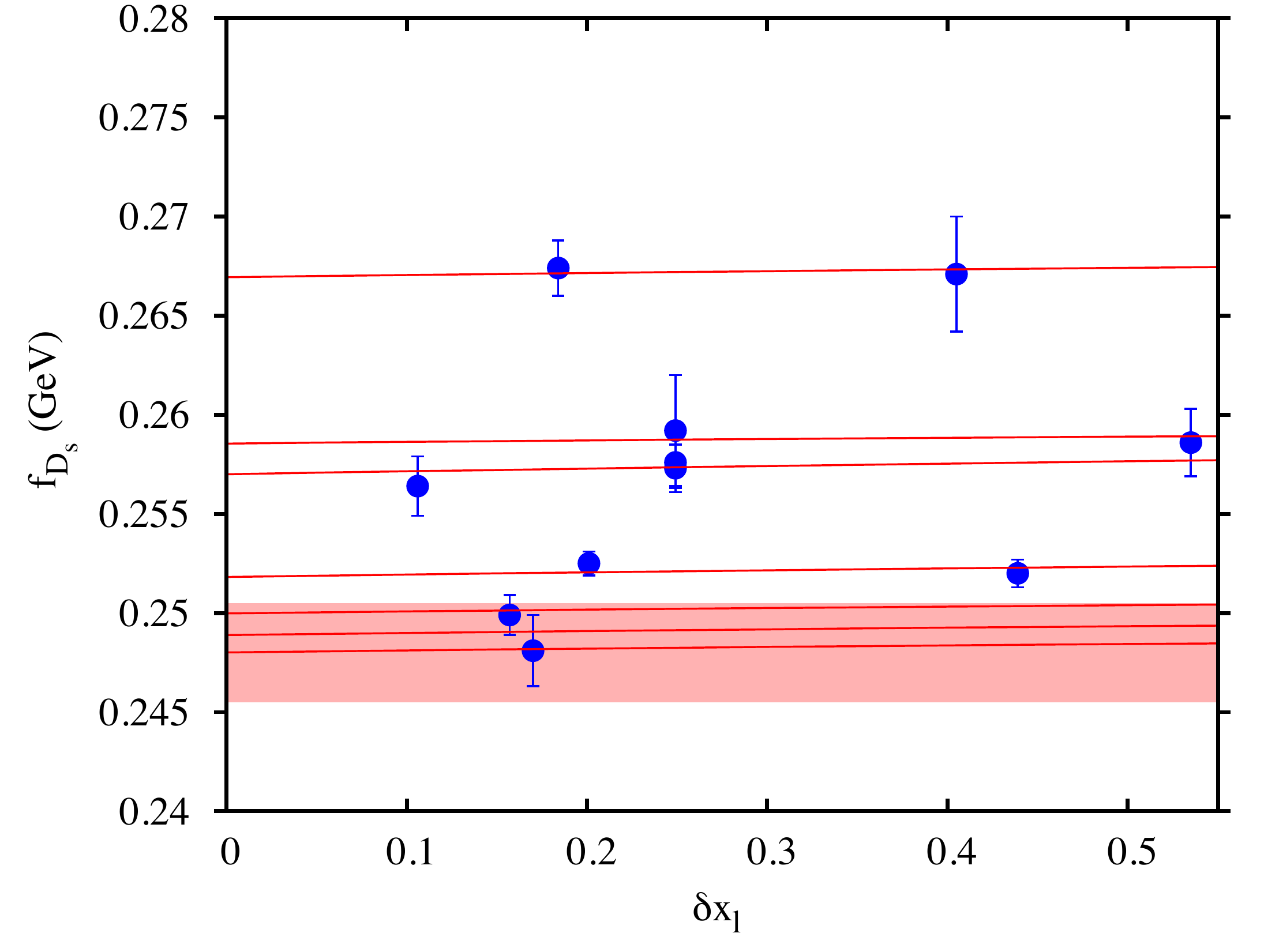}
\end{center}
\caption{Results for the decay constant of the $D_s$ meson 
tuned to the correct valence $c$ and $s$ mass on each 
ensemble from Table~\ref{tab:tune} as a function of the difference 
between the sea light quark mass and the physical value scaled 
by the physical strange quark mass (i.e. the parameter $\delta x_l$). 
The results are clearly separated by their lattice spacing value
with very coarse at the top and ultrafine at the bottom. 
The lines 
show the result of the fit 
described by equation~\ref{eq:fitformf}, taken at the 
value of the sea strange quark mass ($\delta x_s$) and using the 
lattice spacing value corresponding to the ensemble with 
smallest $\delta x_l$ in that group. 
The results on the coarse lattices at $\delta x_l = 0.25$ 
include numbers at two different values of $\delta x_s$ as 
well as at two different volumes. This gives an idea of 
the spread in results from these effects. 
The lowest line is the fit curve in $\delta x_l$ at $a=0$
and $\delta x_s = 0$. 
The shaded red band 
gives our final result 
with the full error as described in the text. 
}
\label{fig:fdsvssea}
\end{figure}

Figure~\ref{fig:fdsvsasq} shows the results plotted against the 
square of the lattice spacing. The line is the fit curve for 
the physical sea quark mass values (i.e. $\delta x_l = \delta x_s = 0$). 
The shaded band is then the final physical result including the 
full error of 1.0\% (2.5 MeV), to be discussed below and broken down into 
its component parts in Table~\ref{tab:errors}. 

We construct the error budget as before, separating the 
error of 1.9 MeV resulting from the extrapolation to 
the physical point into its components of statistical error, 
$r_1/a$ error and errors from extrapolation in the lattice spacing and 
in the sea quark masses. Here the contributions from statistical errors and 
the different extrapolation errors are comparable. 

The error in the physical value of $r_1$ is 0.7\%. This 
becomes a 0.6\% error in $f_{D_s}$ when the effects of 
$r_1$ on shifting the value of $m_{\eta_s}$ are taken into
account. The effect of the 0.6\% uncertainty in the physical value 
of $m_{\eta_s}$ can similarly be estimated from the 
dependence of $f_{D_s}$ on the $\eta_s$ mass at 0.1\%.   
The uncertainty in $f_{D_s}$ from the uncertainty in 
the value of the $\eta_c$ mass is negligible. 
The error from working on a finite spatial volume instead 
of infinite volume is estimated at 0.1\% from comparing 
finite and infinite volume chiral perturbation theory. 
It is clear from our results (see Table~\ref{tab:ds}) that 
we see no significant volume dependence within our
0.5\% statistical errors, which is in agreement with 
chiral perturbation theory, but that provides a stronger 
constraint. 

The size of electromagnetic effects inside the $D_s$ can be bounded 
by the size of these effects on the $\eta_c$. By allowing for an 
electromagnetic contribution to the heavy quark potential we estimate 
that $f_{\eta_c}$ could be increased by up to 0.4\% by these 
effects. Since the $D_s$ has one quark of half the electromagnetic 
charge and is also much larger, so less sensitive to 
short-distance electromagnetic effects, 
we conservatively take an error of 0.1\% from internal 
electromagnetic effects~\cite{footem}. 

The error resulting from missing $c$ quarks in the sea can also 
be bounded by the size of such effects on $f_{\eta_c}$. 
In section~\ref{sec:mds} we discussed a comparison between the hyperfine potential 
in charmonium and that induced by adding $c$ quarks in the 
sea. The hyperfine potential causes the difference between 
$f_{J/\psi}$ and $f_{\eta_c}$, which we will see in the next 
section is very small, 3\%.  The $c$-in-the-sea potential is 
280 times smaller and so will produce a completely negligible effect
on $f_{\eta_c}$ and therefore also on $f_{D_s}$.   

Figure~\ref{fig:fdsvssea} shows the results for $f_{D_s}$ 
as a function of the sea light quark mass, normalised 
to the strange mass as in equation~\ref{eq:delta}. The 
lines show the fitted curves at the appropriate values 
of lattice spacing and sea strange quark mass, along 
with the final physical curve and final result with 
error band. No significant dependence on sea quark 
masses is seen.  

Our final result for $f_{D_s}$ is 0.2480(25) GeV, to be 
compared to the October 2010 average from the Heavy Flavor 
Averaging Group of 0.2573(53) GeV~\cite{hfag}. 

\subsection{$f_{\eta_c}$}
\label{sec:fetac}

Here we study the remaining independent quantity that can 
be extracted from the pseudoscalar correlators calculated 
here, the decay constant of the $\eta_c$ meson. Although 
this cannot be directly related to any process measurable 
in experiment, it can be compared between lattice 
QCD calculations using different formalisms for the 
$c$ quarks. Since we have particularly accurate results 
here, we give a value for $f_{\eta_c}$ that others 
can use to test their formalisms against.  

\begin{figure}
\begin{center}
\includegraphics[width=80mm]{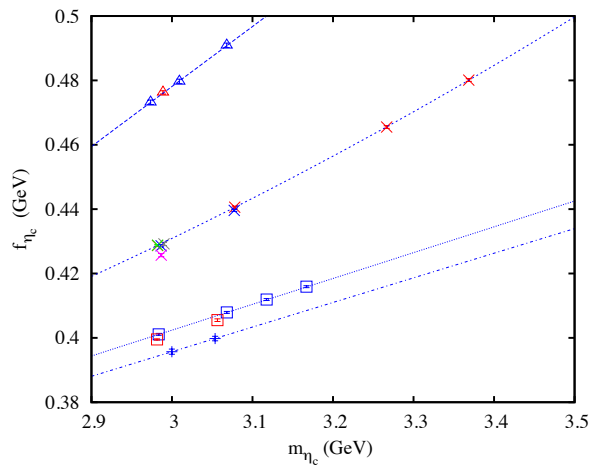}
\end{center}
\caption{Results for the $\eta_c$ decay constant 
as a function of the $\eta_c$ mass for the different ensembles 
in Table~\ref{tab:params}. 
As in Fig.~\ref{fig:fmhhsea}, very coarse 
ensembles are triangles; coarse, crosses; fine, squares; superfine, pluses. 
Errors shown are statistical only. 
Results for heavier sea $u/d$ quark masses at 
each lattice spacing are in 
red, lighter ones are in blue. 
On the coarse lattices the very heavy sea masses of set 6 
are in pink, the lighter strange sea mass of set 7 in grey 
and the large volume results on set 5 are in green, 
on top of the result from set 4. 
The lines are fits to the results for one ensemble at each 
lattice spacing allowing 
linear and quadratic terms in $m_{\eta_c}$. }
\label{fig:fetactunec}
\end{figure}

The raw results for $f_{\eta_c}$ on each ensemble are 
given in Table~\ref{tab:charmmass}. Since the $\eta_c$ 
contains only charm quarks we have only to plot 
$f_{\eta_c}$ against $m_{\eta_c}$ to interpolate to the 
correct point on each ensemble. 
Because this is simpler than having to separate the 
dependence on two masses, as was done for the $D_s$, 
we can plot the results from many more of the 
ensembles. Figure~\ref{fig:fetactunec} shows the results.
As expected, the dependence is linear (we allowed for 
quadratic terms in the fit, but these were small) but with a slope 
that depends on the lattice spacing. The figure also 
emphasises how little sea quark mass dependence there is, 
in line with the evidence from Figure~\ref{fig:fmhhsea}.
Some is visible above our very small statistical 
errors on the coarse and fine ensembles, however. 

Again we use the dependence shown in Figure~\ref{fig:fetactunec} 
to make small tuning shifts 
to the values of $f_{\eta_c}$ on each ensemble so that they 
correspond to the correct result for $m_{\eta_c}$ = 2.985 GeV. 
Table~\ref{tab:tune} gives these tuned values.  The statistical/tuning 
errors are small but the $r_1/a$ errors are even smaller because 
of cancellation when the retuning is done on changing the 
lattice spacing. Once again the lattice spacing dependence 
is the most striking feature of these results. 

We fit the tuned values to the same functional form as 
used for $m_{D_s}$ (equation~\ref{eq:fitform}) 
and $f_{D_s}$ (equation~\ref{eq:fitformf}).    
We take the same prior values and widths for the parameters 
except that for the physical value of $f_{\eta_c}$, $f_{\eta_c, phys}$ 
we take 0.4(2) and for the coefficients, $c_i$, for the 
discretisation errors we take 0.0(3), since $v^2$ for a 
$c$ quark is expected to be somewhat higher than in a $D_s$. 

\begin{figure}
\begin{center}
\includegraphics[width=80mm]{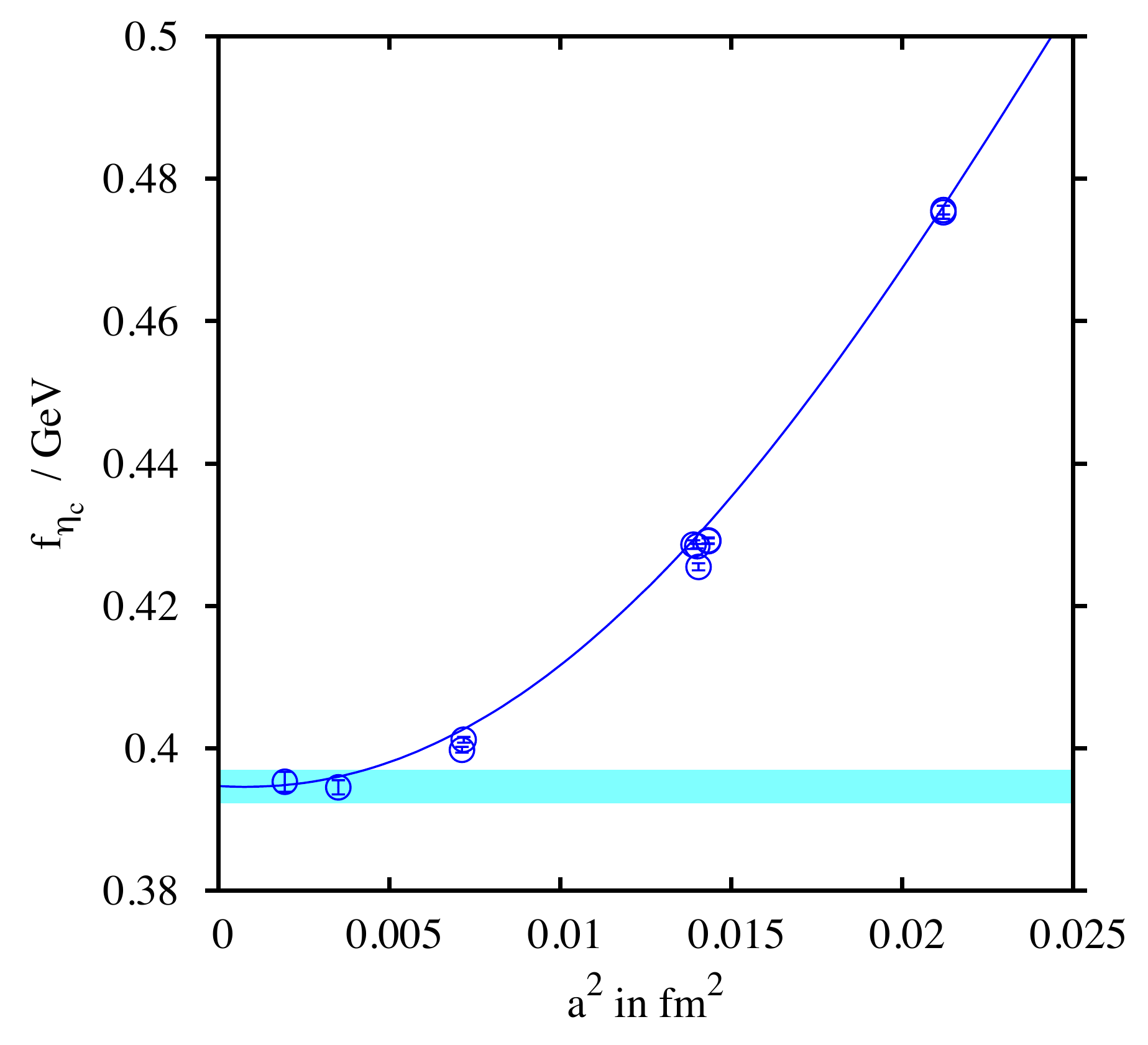}
\end{center}
\caption{Results for the $\eta_c$ decay constant
tuned to the correct $c$ mass on each 
ensemble as a function of the square of the lattice 
spacing. The line shows the result of the fit 
described in the text.  
The shaded band gives our final result with the 
full error bar as described in the text. 
}
\label{fig:fhhasq}
\end{figure}

\begin{figure}
\begin{center}
\includegraphics[width=80mm]{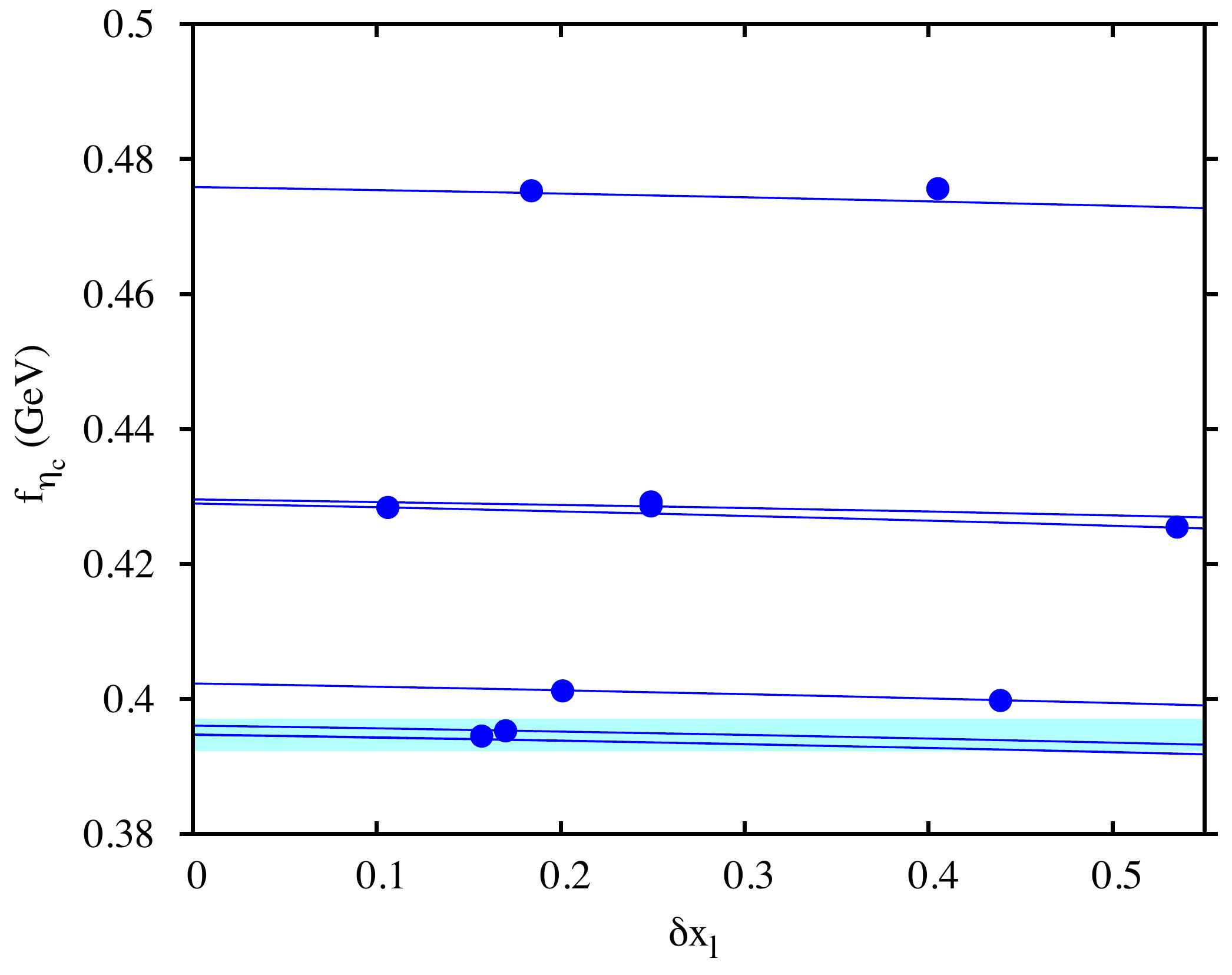}
\end{center}
\caption{Results for the decay constant of the $\eta_c$ meson 
tuned to the correct valence $c$ mass on each 
ensemble from Table~\ref{tab:tune} as a function of the difference 
between the sea light quark mass and the physical value scaled 
by the physical strange quark mass (i.e. the parameter $\delta x_l$). 
The results are clearly separated by their lattice spacing value
with very coarse at the top and ultrafine at the bottom. 
The lines 
show the result of the fit 
described in the text, taken at the 
value of the sea strange quark mass ($\delta x_s$) and using the 
lattice spacing value corresponding to the ensemble with 
smallest $\delta x_l$ in that group. 
The results on the coarse lattices at $\delta x_l = 0.25$ 
include numbers at two different values of $\delta x_s$ as 
well as at two different volumes. This gives an idea of 
the spread in results from these effects. 
The lowest line is the fit curve in $\delta x_l$ at $a=0$
and $\delta x_s = 0$. 
The shaded blue band 
gives our final result 
with the full error as described in the text. 
}
\label{fig:fetacvssea}
\end{figure}

The extrapolated value at the physical point, $f_{\eta_c, phys}$,
 is 0.3947(20) GeV with a $\chi^2/{\rm dof}$ of 0.3 for 11 
degrees of freedom. Once again we tested how robust the fit was:
\begin{itemize}
\item changing the prior on all the $c_i$ (including $c_1$) 
to 0.0(8) changes $f_{\eta_c, phys}$ by 0.5$\sigma$ (1 MeV) 
and increases the error by 40\%. 
\item adding two powers of $a^2$ into the sum on $j$ in 
the fit equation does not change $f_{\eta_c,phys}$ or its error; 
subtracting two powers changes $f_{\eta_c, phys}$ by 0.5$\sigma$ (1MeV) 
and reduces the error by 30\%.  
\item adding an extra power of discretisation errors into both the 
linear and quadratic sea-quark mass 
dependent terms makes no difference. 
\item  missing out the sea quark mass dependence altogether 
does not change $f_{\eta_c, phys}$ but increases the
$\chi^2$ value to 1. 
\item Changing all the $\delta x$ values 
by 10\% in either direction makes no appreciable difference, nor does 
changing them within their error bars on, for example, the 
ultrafine or fine lattices. 
\item missing out the very coarse lattice results does not 
change $f_{\eta_c, phys}$ appreciably; 
neither does missing out the very coarse and the coarse but 
the error increases by 50\%. 
\item missing out the ultrafine result shifts $f_{\eta_c, phys}$ by 1.4$\sigma$ 
(2.5 MeV) and increases the error by 40\%.  
\end{itemize} 

The error budget is constructed as before, estimating 
the split in the error obtained from the fit into 
components from statistics, $r_1/a$ and extrapolations 
in $a^2$ and the sea quark masses. In addition the 
error from the uncertainty in the physical value of 
$r_1$ becomes 0.3\%, allowing for the cancellation 
that reduces the sensitivity below the naive 0.7\%. 
The error from finite volume effects we take to be 
negligible based on the chiral perturbation theory 
studies of the much larger $D_s$ meson. 

As we will discuss in section~\ref{sec:disc}, $f_{\eta_c}$ 
is not a quantity that can be compared directly 
to experiment. We include it here as a calibration 
point for lattice QCD studies of charm physics. 
As such, we do not have to include errors arising 
from effects outside a pure lattice QCD calculation 
including $u$, $d$, and $s$ sea quarks and taking 
the $\eta_c$ mass to be 2.985 GeV. Thus in Table~\ref{tab:errors} 
only the top six errors in the final column 
should be included for such a calculation 
and the bottom three ignored. 

For completeness we discuss other sources of error 
that may need to be considered if lattice QCD calculations 
differing in detail from ours are compared to it. 
The error that arises from the 3 MeV uncertainty in the 
physical value of the $\eta_c$ mass can be estimated 
from the slope of $f_{\eta_c}$ with $m_{\eta_c}$ in 
Fig~\ref{fig:fetactunec}. This gives an error of 0.05\% with 
$f_{\eta_c}$ increasing with the value of $m_{\eta_c}$. 
Internal electromagnetic effects inside the $\eta_c$ 
will also increase $f_{\eta_c}$. In section~\ref{sec:fds}
we estimated this effect at 0.4\% (but lattice QCD
calculations will not typically include electromagnetic effects). 
The effect of including $c$ quarks in the sea will 
also be to increase $f_{\eta_c}$. In section~\ref{sec:fds} 
we estimated this as 0.01\%, based on a comparison to 
$f_{J/\psi}$ that will be described in section~\ref{sec:discetac}. 

Figure~\ref{fig:fhhasq} shows $f_{\eta_c}$ against 
$a^2$ in ${\rm fm}^2$ with the fit curve for 
the physical sea quark mass values. The shaded band 
is the final physical result including the full 0.6\% 
error i.e. 0.3947(24) GeV. 

Figure~\ref{fig:fetacvssea} shows the results for $f_{\eta_c}$ 
as a function of the sea light quark mass, normalised 
to the strange mass as in equation~\ref{eq:delta}. The 
lines show the fitted curves at the appropriate values 
of lattice spacing and sea strange quark mass, along 
with the final physical curve and final result with 
error band. No significant dependence on sea quark 
masses is seen.  

\section{Discussion}
\label{sec:disc}

A summary of the results from this calculation is 
then:
\begin{eqnarray}
\label{eq:res}
m_{D_s} &=& 1.9691(32) GeV \\ \nonumber
f_{D_s} &=& 0.2480(25) GeV \\ \nonumber
f_{\eta_c} &=& 0.3947(24) GeV
\end{eqnarray}

\subsection{Comparison to our previous results}

Our new results improve on our 2007 results~\cite{fds} in several 
ways, as described earlier.  It is worth discussing 
the effect of these changes on the final numbers 
because, particularly in the case of $f_{D_s}$, the 
shift from 2007 is significant. 

Our 2007 result for $m_{D_s}$ was 1.962(6) GeV obtained 
from very coarse, coarse and fine ensembles. The lattice 
spacing was fixed using the quantity $r_1$ as here, but 
setting the physical value of $r_1$ to 0.321(5) fm. 
The error on $m_{D_s}$ from this uncertainty in $r_1$ was 
0.2\% i.e. 4 MeV. Since then we have improved significantly 
the calibration of the lattice spacing by improving 
the determination of the physical value of $r_1$ 
to 0.3133(23) fm. This has used improved determinations 
of $r_1/a$ on each ensemble by the MILC collaboration~\cite{milcreview}.   
The change in the value of $r_1$ represents 1.5$\sigma$ 
and therefore we expect $m_{D_s}$ to change by approximately 
6 MeV. In fact the change has been 7 MeV. 

Of course the change in $r_1$ has not been the only change. 
The lattice spacing values on individual ensembles have 
moved relative each other with changes in $r_1/a$ values. 
These have moved furthest on the very coarse set 1, 
changing by 1\% or 2$\sigma$, but with some changes of up to 0.5\%
(1$\sigma$) on the coarse ensembles. Values on the fine ensembles have 
not shifted signficantly. The relative shifts change 
the lattice spacing extrapolation slightly, as does our 
improved tuning of the charm quark mass (the strange mass 
was tuned within the chiral extrapolation previously 
using results for the $K$ and $\pi$ meson masses). We also 
have 
additional determinations of the sea quark mass dependence. 
These other effects largely cancel each other, however, 
in this case. Our new error budget shows an improved 
error coming from the determination of $r_1$ and this is 
the main effect behind the reduction of total error 
from 0.3\% to 0.2\%. 

Our $f_{D_s}$ result has changed by 2.7\% (from 0.2415(32) GeV) 
which is a shift of 2$\sigma$. From our error budget the change
expected from the change in $r_1$ is 1.5\%. Combined with changes 
in $r_1/a$ and improved tuning, however, results on 
the fine and very coarse ensembles have changed 
by up to 2\%. This has affected the continuum extrapolation.
Sea quark mass effects, although not significant either now or 
before, have also changed in the same direction. This has 
meant that the 0.3\% sea quark mass extrapolation error 
has added linearly to (some of) the 0.5\% continuum extrapolation 
and the roughly 2\% shift, rather than in quadrature. 

The ratio $f_{D_s}/f_D$ is not very sensitive to $r_1$ and 
so, although we have not yet performed an improved analysis 
of $f_D$, we would not expect this ratio to change very much. 
If we take our previous result for $f_{D_s}/f_D$, but double 
the $r_1$ uncertainty and add it linearly to the 
 $a^2$ and $m_{u,d}$ extrapolation errors to allow 
for the behaviour seen in $f_{D_s}$ we would obtain an 
error of 1.5\% on the ratio, giving 1.164(18). 
Combined with our new result for $f_{D_s}$ this 
gives a value for $f_D$ of 0.213(4) GeV, to be compared 
to a CLEO result of 0.206(9) GeV~\cite{cleofd}. We emphasise that 
our new value for $f_D$ does not result from a new 
analysis of $f_D$ itself but simply from the change 
in $f_{D_s}$ given here.  

Our 2007 results for $f_\pi$ and $f_K$ change a little when 
the new value for $r_1$ is used. Using the fitting procedure 
described in the appendices of \cite{r1paper} 
(but not including the experimental results for 
$f_\pi$ and $f_K$ in the fit data), we find 
$f_\pi=132(2)$\,MeV and $f_K=159(2)$\,MeV which 
agree within errors with our 2007 results~\cite{fds} and 
with experiment~\cite{pdg} to within about $1.5\sigma$. 

\subsection{Comparison to other lattice results and to experiment}

\subsubsection{$m_{D_s}$}

As discussed in section~\ref{sec:mds}, the accurate determination 
of the mass of the $D_s$ meson is an important test of the 
calculation of $f_{D_s}$. Our result, 1.9691(32) GeV, is in 
good agreement with experiment, as shown in Figure~\ref{fig:mdsvsasq}.
The experimental error is 0.3 MeV~\cite{pdg}. To improve the lattice 
QCD error of 3 MeV further would require improved statistical 
errors on the very fine lattices but also improved errors 
from electromagnetic/$\eta_c$ annihilation effects that 
are not currently included in lattice QCD calculations.  
It is impressive that lattice QCD calculations have reached 
the point where electromagnetic effects have to be considered 
in the match to experiment. 

\begin{figure}
\begin{center}
\includegraphics[width=80mm]{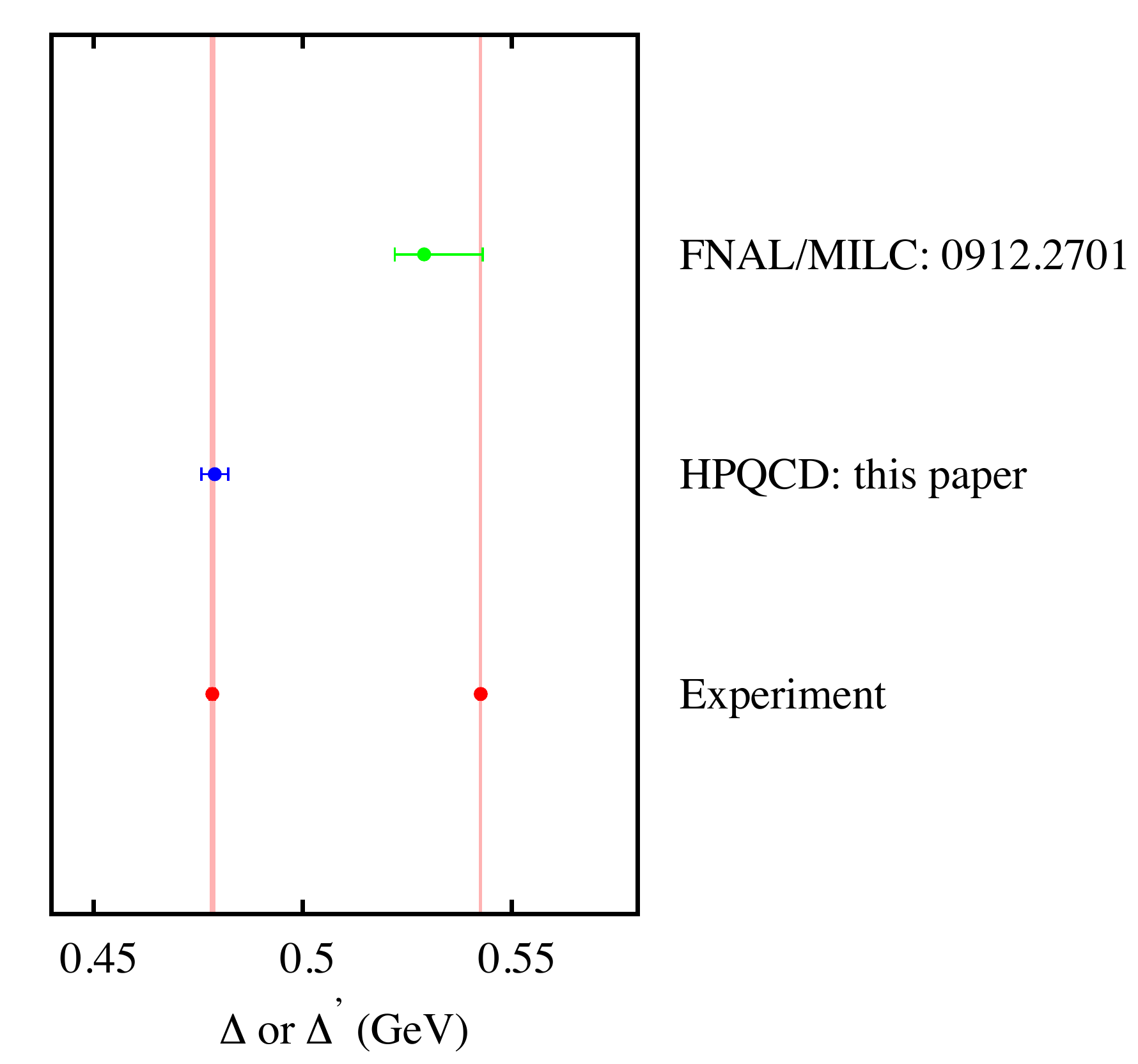}
\end{center}
\caption{Summary of full lattice QCD results for the difference 
of binding energy between charm-strange and charmonium 
states. $\Delta$ uses pseudoscalar mesons $D_s$ and 
$\eta_c$ and compares the result from this paper to experiment, given 
by the appropriate red point and shaded band). $\Delta^{\prime}$
uses a spin-average of the pseudoscalar and vector states  
and compares the result from the Fermilab Lattice/MILC 
collaborations to experiment. 
Our result corresponds to the complete error budget given in 
Table~\ref{tab:errors} and is corrected for missing electromagnetic 
effects. The Fermilab Lattice/MILC result includes both 
errors given in~\cite{fnaltune} but has not been corrected for 
missing electromagnetic effects. 
}
\label{fig:mdssummary}
\end{figure}

Other lattice QCD formalisms for $c$ quarks are not as highly 
improved as HISQ. They then have more difficulty in handling 
charmonium and so fix the $c$ quark mass from the $D_s$. 
However, we believe that it is still important to check the 
masses of other mesons containing $c$ quarks as a test of 
systematic errors. The easiest quantity to compare
is the one defined earlier as $\Delta = m_{D_s}-m_{\eta_c}/2$, the difference in 
binding energy between charmonium and $D_s$. 
Our result for this is plotted in Figure~\ref{fig:mdssummary}. 
A variant of $\Delta$ was recently calculated using the Fermilab 
heavy quark formalism for $c$ quarks, combining this with light 
asqtad quarks on the MILC very coarse, coarse and fine 
ensembles~\cite{fnalcharm}. The $c$ mass is fixed from the 
energy-momentum relation for the $D_s$ meson (because 
the energy at zero momentum is not equal to the mass), which leads 
to sizeable statistical errors in the tuning process, 
growing with heavy quark mass~\cite{fnaltune}. 
Typically the `kinetic mass' for the $D_s$ is 
obtained to 2\%.  
The Fermilab lattice/MILC collaborations 
quote a result for 
$\Delta^{\prime} = m(\overline{D_s}) - m(\overline{1S})/2$ 
of 0.529$\pm 7 {+12 \atop -0} $ with a partial error budget~\cite{fnalcharm}. 
Here $\overline{D_s}$ indicates the spin-average mass of 
the $D_s$ and the $D_s^*$ and $m(\overline{1S})$ is the 
spin average of the masses of the $J/\psi$ and the $\eta_c$. 
The spin average is used to reduce their discretisation 
error from spin-dependent terms, but the $D_s^*$ does have 
a strong decay mode, albeit Zweig-suppressed, that will 
lead to an additional systematic error in the lattice QCD
calculation.   
The first error given above is from statistics and extrapolation 
uncertainties and the second from the physical value of 
$r_1$ which they take as $0.318 {+0.000 \atop -0.007}$ fm. The Fermilab Lattice/MILC result agrees with experiment and 
is plotted in Figure~\ref{fig:mdssummary} for 
comparison to our result for $\Delta$. More detailed 
comparison between the results needs improved accuracy 
for those from the Fermilab formalism. 

\subsubsection{$f_{D_s}$}

\begin{figure}
\begin{center}
\includegraphics[width=80mm]{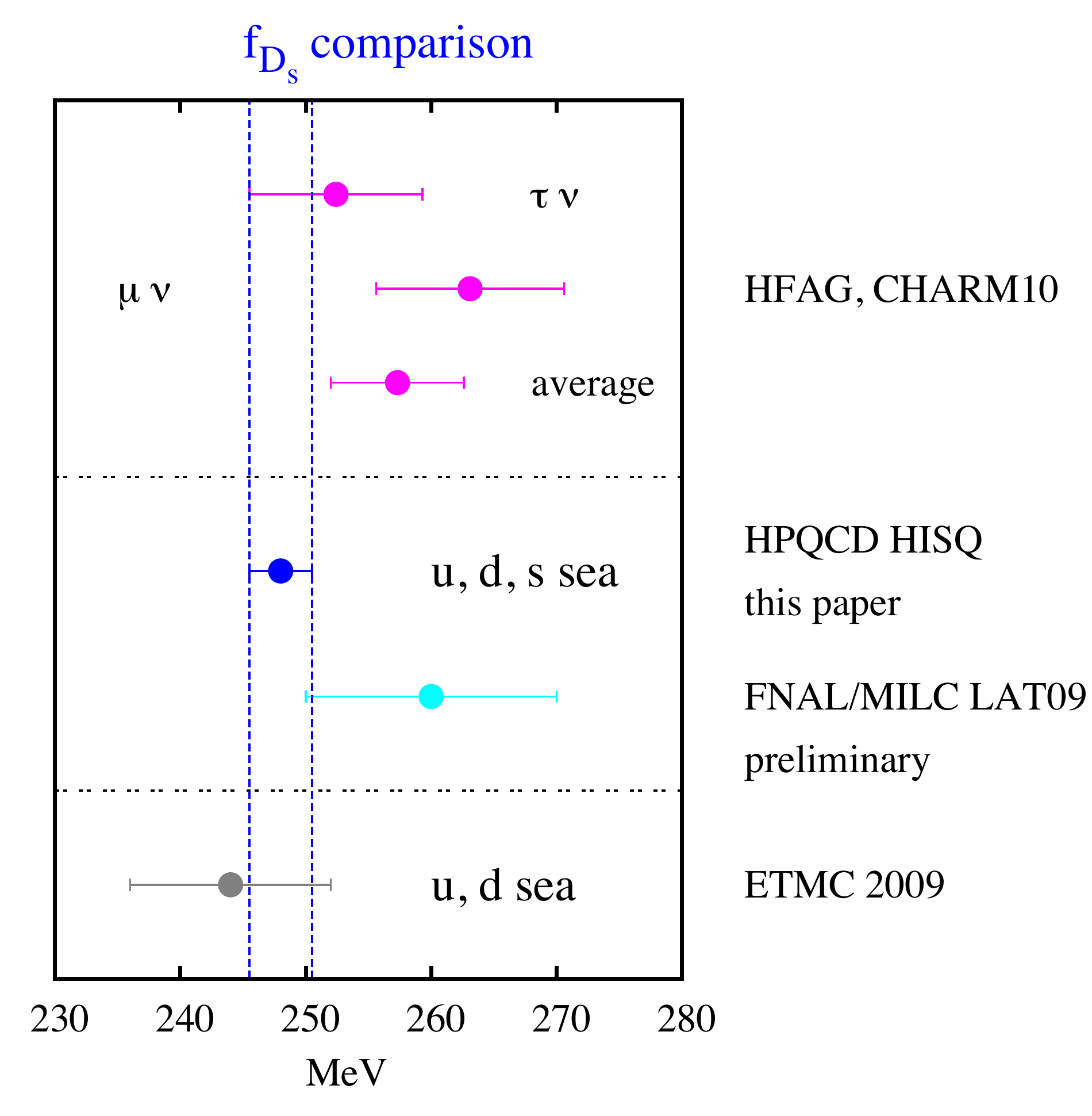}
\end{center}
\caption{Comparison of the result from this paper 
for the $D_s$ decay 
constant with those from other lattice QCD calculations 
that include the effect of sea quarks. 
The Fermilab Lattice/MILC result is a preliminary 
one but also includes the effects of $u$, $d$ and 
$s$ quarks in the sea. The ETMC result includes 
only $u$ and $d$ quarks in the sea. 
We show also a recent average of experimental results
from the Heavy Flavor Averaging Group~\cite{hfag} 
and two separate averages over the $\mu \nu$ and 
$\tau \nu$ channels. 
Experimental results for $f_{D_s}$ convert the 
leptonic decay rate to a 
decay constant using equation~\ref{eq:fdsexpt} and 
using an input value for $V_{cs}$ (see text). 
}
\label{fig:fdscomp}
\end{figure}

\begin{figure}
\begin{center}
\includegraphics[width=80mm]{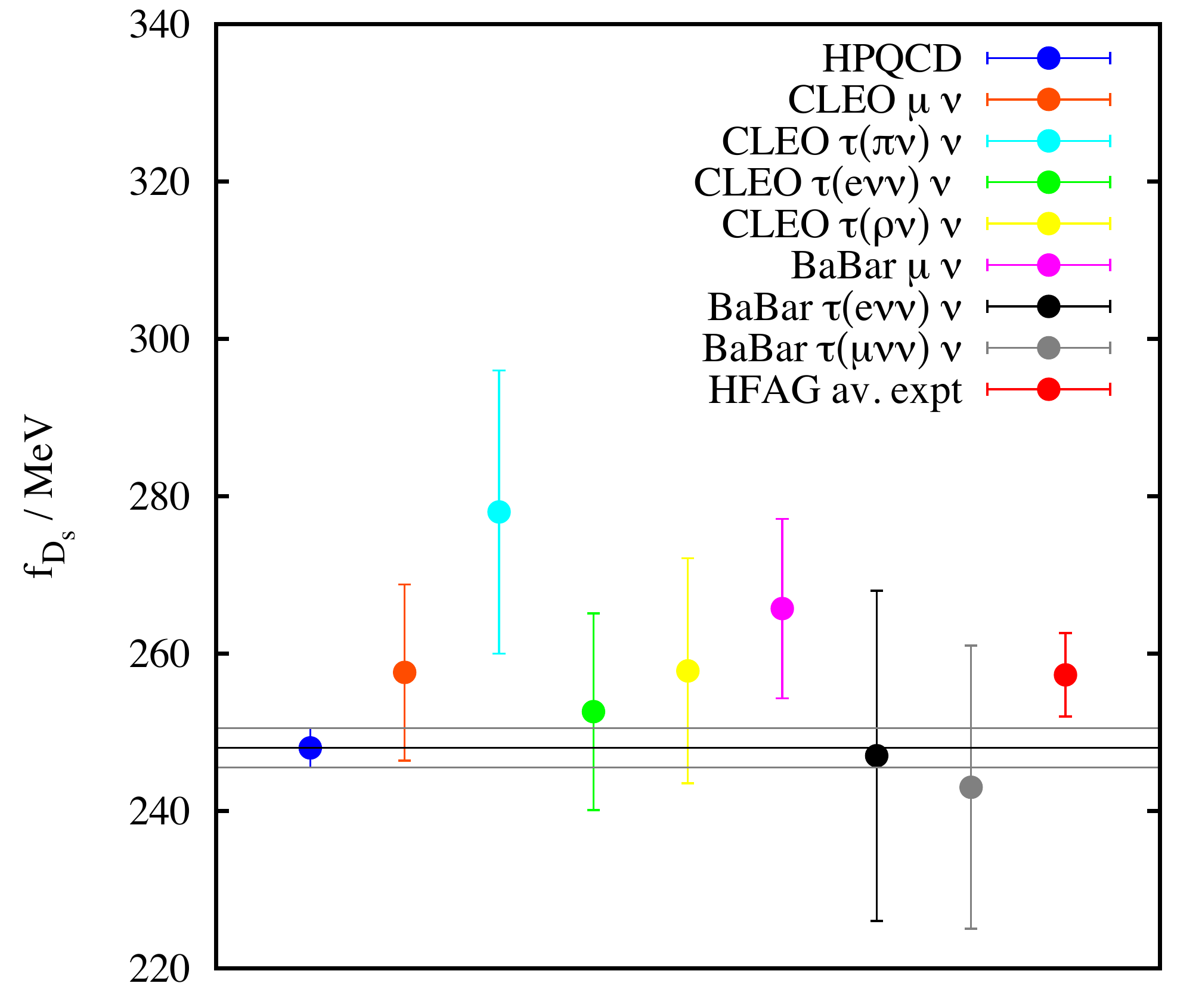}
\end{center}
\caption{Comparison of our new result for the $D_s$ decay 
constant with recent experimental results from CLEO~\cite{cleofdsmu, cleofdste, cleofdstrho} 
and BaBar~\cite{babar10new}. These 
are derived from leptonic decay modes of the $D_s$ 
in various channels, and using equation~\ref{eq:fdsexpt} with 
an input value for $V_{cs}$ (see text). The CLEO numbers are 
taken from the compilation in~\cite{cleofdstrho}, 
using consistent values for $V_{cs}$, $m_{D_s}$ and $\tau_{D_s}$ and so 
differ slightly from the historical numbers in Figure~\ref{fig:fdshistory}.  
We also include the HFAG 2010 world average for experiment~\cite{hfag}. 
}
\label{fig:exptcomp}
\end{figure}

Figure~\ref{fig:fdscomp} compares the result for the 
$D_s$ decay constant from this paper to other 
lattice QCD calculations that include the effect of 
sea quarks. The Fermilab Lattice/MILC result of 260(10) MeV is 
a preliminary one from a conference presentation~\cite{fnallat09}, updated
from their original 2005 calculation~\cite{fnalfd} in a number of 
ways but including an update of the physical value of 
the parameter $r_1$ used to set the lattice spacing as
we have done here. Their calculation uses MILC 
gluon field configurations as we do, but at the 
three coarsest lattice spacing values that we have used. 
The Fermilab formalism for $c$ quarks is combined with 
the asqtad formalism for the $s$ quarks. As explained 
above the $c$ quark mass is tuned from
the dispersion relation for $D_s$ mesons. In the Fermilab 
formalism there is no PCAC relation and so the 
temporal axial current operator that annihilates the 
$D_s$ in its leptonic decay (equation~\ref{eq:fdef}) 
must be renormalised 
to match the continuum current operator that couples 
to the $W$. This is done by a perturbative calculation 
to $\cal{O}$$(\alpha_s)$ after taking a ratio to 
vector current operators. The systematic uncertainty from 
this approach is in principle $\cal{O}$$(\alpha_s^2)$ 
($\approx$ 5\%), but it is argued in~\cite{fnallat09, fnalpert} 
that a significantly smaller (1.4\% + 0.3\%)
error be used which is the square of the one-loop 
contribution. It would be useful to test this on 
a calculation such as $f_K$ where the result is well-known~\cite{craigfk}.   
With relativistic formalisms such as the HISQ formalism 
used here and the twisted mass formalism to be discussed 
below, the existence of the PCAC relation means that 
the issue of renormalisation does not arise. Also 
in both cases, $f_K$ can be calculated as well 
as $f_{D_s}$ as a test of the error analysis. 

Figure~\ref{fig:fdscomp} also includes the result 
244(8) MeV
from the European Twisted Mass Collaboration~\cite{etmcfds} using 
the twisted mass formalism for all of the quarks. 
This formalism is an improved version of the Wilson 
formalism with discretisation errors starting at 
$\cal{O}$$(a^2)$, somewhat worse than the $\cal{O}$$(\alpha_s a^2)$ 
for HISQ, but also having a partially conserved axial 
current so no renormalisation issues. 
ETMC include only the effect of $u$ and $d$ quarks 
in the sea, however, and it is not clear what 
systematic error to take for missing $s$ quarks that 
are there in the real world. We cannot 
use perturbative arguments, as we have done here to 
account for the missing $c$ quarks in the sea. 
ETMC are now improving their calculations to include 
both $s$ and $c$ sea quarks~\cite{urbach}. 

The experimental results shown on figure~\ref{fig:fdscomp} 
are the October 2010 averages from the Heavy Flavor Averaging Group~\cite{hfag}, 
using recent CLEO~\cite{cleofdsmu, cleofdste, cleofdstrho}, BaBar~\cite{babar10, babar10new} and Belle~\cite{bellefds} results from 
measurement of the $D_s \rightarrow \mu \nu$ and $D_s \rightarrow 
\tau \nu$ decay rates. 
To determine $f_{D_s}$ from experiment the measured leptonic 
branching fraction, corrected for electromagnetic radiation~\cite{footem},
is used in:
\begin{equation}
f_{D_s} = \frac{1}{G_F |V_{cs}| m_l (1-m_l^2/m_{D_s}^2)}\sqrt{\frac{8\pi{\cal{B}}(D_s \rightarrow l \nu)}{m_{D_s}\tau_{D_s}}}.
\label{eq:fdsexpt}
\end{equation}
A value for $V_{cs}$ must be assumed. In the 
past $V_{cs} = V_{ud}$ has often been taken (see, 
for example,~\cite{cleofdsmu}), assuming $2\times 2$ CKM unitarity. 
HFAG take the 2010 Particle Data Tables result 
for $V_{cs}$ (0.97345(16)) from a full CKM matrix unitarity fit~\cite{hfag, pdg}. 
These two alternatives for $V_{cs}$ differ at the level 
of 0.1\% which is irrelevant here. 

It is clear from Figure~\ref{fig:fdscomp} that there is no longer any 
significant `$f_{D_s}$ puzzle'~\cite{kronfeld09} since 
the discrepancy between our lattice 
QCD result and the world average of experiment (257.3(5.3) MeV)
 is $1.6\sigma$. 
The average of experimental results in the $\tau \nu$ channel 
(252.4(6.9) MeV) and our 
value agree very well. 
This is emphasised further in Figure~\ref{fig:exptcomp} where 
the most accurate recent experimental results are 
individually compared to our value for $f_{D_s}$, and all except one
disagree by less than $1\sigma$.  

\begin{figure}
\begin{center}
\includegraphics[width=80mm]{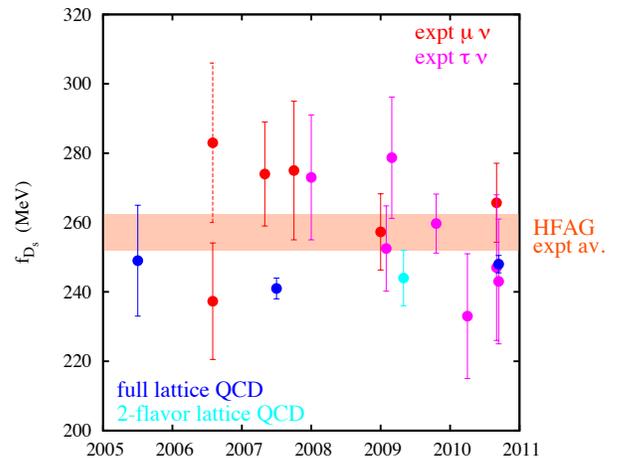}
\end{center}
\caption{ Values for $f_{D_s}$ from experiment 
and from lattice QCD since 2005, excluding results 
from conference proceedings. Later results from a 
given collaboration and process supersede the earlier 
ones. Experimental results 
are divided into those from the $\mu \nu$ channel~\cite{babar07fds, cleo07fds, bellefds, cleofdsmu, babar10new} 
(in red)
and those from the $\tau \nu$ channel~\cite{cleo07fds, cleofds08, cleofdsmu, cleofdste, cleofdstrho, babar10, babar10new} (in several 
$\tau$ decay modes, in pink).  
The HFAG October 2010 world average for 
experiment~\cite{hfag} is included 
as a light orange band. Note that the leftmost red point 
(from BaBar~\cite{babar07fds}) appears with dashed error bars - the lower 
value with solid error bars is the result adjusted by 
HFAG~\cite{hfagnote}, 
although 
this number is not now included in the HFAG average. 
Lattice QCD points are in dark blue for full QCD~\cite{fnalfd, fds} - the 
rightmost point is from this paper. The light blue 
point is from ETMC~\cite{etmcfds} including only $u$ and $d$ in the sea.  
}
\label{fig:fdshistory}
\end{figure}

Things have now changed quite significantly 
since the summer of 2008 when the most accurate experimental 
result for $f_{D_s}$ was 267.9(9.1) MeV~\cite{cleo-ichep08} and the most 
accurate lattice QCD result was 
241(3) MeV~\cite{fds, me-ichep08}, differing by 
almost 3 $\sigma$. The experimental average moved 
down 5\% (1.5$\sigma$) in early 2010 but has since moved
up 1\% to the new world average value
and the lattice result has moved up 3\% (2.3$\sigma$). 
The discrepancy between experiment and lattice QCD is 
now only 4\% (1.6 $\sigma$) and the experimental error 
is now reduced to only twice that of the lattice 
QCD error. This marks significant effort both 
experimentally and theoretically on this quantity 
to understand and pin down the original discrepancy. 
Figure~\ref{fig:fdshistory} shows the history 
of $f_{D_s}$ from experiment and lattice QCD since 
the first full lattice QCD calculation of 2005.

\subsubsection{$f_{\eta_c}$}
\label{sec:discetac}

As stated earlier, there is no direct comparison 
possible between lattice results for $f_{\eta_c}$ and 
experiment because the $\eta_c$ does not annihilate 
to a $W$ boson or other particle that would couple 
directly to the temporal axial current. The high 
accuracy of our results is therefore useful only 
to provide a comparison point for other lattice 
QCD calculations. No result of comparable accuracy 
is available from any other charm quark formalism  
as yet and including the effect of sea quarks.  
ETMC~\cite{etmcfetac} quote a preliminary result of 379(29) MeV 
for $f_{\eta_c}$ including $u$ and $d$ quarks in 
the sea only and tuning the $c$ mass from the 
mass of the $J/\psi$ (i.e. this analysis is 
not directly linked to their $D_s$ analysis, as ours is). 
Future lattice charmonium calculations using different 
formalisms (for example~\cite{trinlat} or~\cite{gunnar}) can use our result 
as a benchmark point to 
check renormalisation or discretisation effects 
because $f_{\eta_c}$ is a very simple quantity 
to calculate. 

Although direct comparisons with experiment do 
not exist, various comparisons that rely on 
approximation schemes, principally potential 
models, can be made.  In a potential model 
the decay constant of an S-wave state is related to the 
wave-function at the origin, $\psi(0)$, by 
$\psi(0) = f\sqrt{M/12}$, where $M$ is the meson 
mass and $f$ its decay constant. 
This relationship is only correct up 
to relativistic and radiative corrections, which for the 
$\eta_c$ could be sizeable (at the level of 30\%) . 
Using this same potential model approach the 
leading term in the decay width for $\eta_c \rightarrow \gamma \gamma$ 
can be written as~\cite{kwongrosner}: 
\begin{equation}
\Gamma(\eta_c \rightarrow \gamma \gamma) = \frac{12 \pi e_c^4 \alpha^2 |\psi(0)|^2}{m_c^2}.
\end{equation}
Here the $c$ quark has electromagnetic charge $e_c$ 
(in units of $e$), mass $m_c$ and
$\alpha$ is the electromagnetic coupling constant. 
This formula has radiative and relativistic corrections 
at the next order. 
The decay width is only poorly known for the $\eta_c$ with the PDG estimate 
given as 7.2(2.1) keV~\cite{pdg}.  
Substituting the decay constant into the formula and taking $m_c = M_{\eta_c}/2$, justifiable 
at this order, gives $f_{\eta_c}$ = 0.4(1) GeV, where only the large error from experiment is 
shown. Alternatively 
one may extract $f_{\eta_c}$ from $B$ decays to 
$\eta_c K$ using the factorization 
approximation. CLEO obtain $f_{\eta_c}$ = 0.335(75) GeV~\cite{cleo-fetac}.  

A more useful experimental result to compare to our decay constant
 is probably to the decay constant of the $J/\psi$. 
Because the $J/\psi$ can annihilate to a photon (seen 
as two leptons in the final state) through the 
vector current there is an exact relationship 
between the decay width and the decay constant 
of the vector particle defined in an 
analogous way to that for the pseudoscalar meson by 
\begin{equation}
\sum_i <0|\overline{\psi}\gamma_i\psi | V_i>/3 = f_V m_V.
\end{equation}
This decay constant can also be calculated in lattice QCD~\cite{hisq2}. Work is in progress and 
results will be given elsewhere.  
The relationship between decay width and decay constant for the process $V_h \rightarrow e^+e^-$ 
is then 
\begin{equation}
\Gamma(V_h \rightarrow e^+e^-) = \frac{4\pi}{3}\alpha_{QED}^2 e_Q^2 \frac{f_V^2}{m_V}
\end{equation} 
The experimental results~\cite{pdg} give $f_{J/\psi}$ = 407(5) MeV 
using $1/\alpha_{QED}(m_c) = 134$~\cite{alpha-em}.  
Thus 1\% accurate 
results for this decay constant are available from experiment, and can be used to test 
lattice QCD.  
In a potential model vector and pseudoscalar values of $\psi(0)$ 
should differ only 
by relativistic corrections, since this is a spin-dependent effect which appears first 
at sub-leading order in the velocity-squared of the heavy quark. Thus 
we would expect our results for the pseudoscalar decay constant to be fairly close
to those for the vector. It is hard to make this statement quantitative however because, even if the 
difference in $\psi(0)$ values of vector and pseudoscalar were accurately pinned down, 
the relationship of $\psi(0)$ to the decay constant 
could have sizeable radiative and relativistic corrections. 

Our result for $f_{\eta_c}$, 0.3947(24) GeV,  is in fact very close to 
the experimental result for $f_{J/\psi}$, only differing by 3\% (2$\sigma$). 
This is somewhat surprising, given naive potential model arguments. 
Accurate lattice QCD studies in bottomonium will show whether this is a coincidence at the charm mass
or a more general feature.

\section{Conclusions}
\label{sec:conc}

In this paper we have updated our 2007 result for the mass 
and decay constant of the $D_s$ meson~\cite{fds} to incorporate a 
new more accurate calibration of the energy scale in 
lattice QCD. We have also included results at two 
finer values of the lattice spacing so we now 
cover a range of lattice spacing values from 0.15 fm 
down to 0.044 fm for improved determination of the 
continuum limit. Our results for $m_{D_s}$ and 
$f_{D_s}$ increase as a result of this calibration. 
$m_{D_s}$ is in excellent agreement with experiment 
with a reduced (3 MeV) error to give 1.9691(32) GeV. 
Our result for $f_{D_s}$ has increased significantly 
to 0.2480(25) GeV. This, along with recent 
movement of the experimental results, means that 
the `$f_{D_s}$ puzzle' is essentially solved: there 
is no longer significant disagreement between theory 
and experiment for this quantity. The experimental 
error is double the theoretical error, however, 
and improved experimental results from BESIII 
aim to obtain a 1\% on $f_{D_s}$~\cite{bes}. The lattice QCD 
error could be further reduced by improved statistical 
accuracy on the very fine lattices.  

\begin{figure}
\begin{center}
\includegraphics[width=80mm]{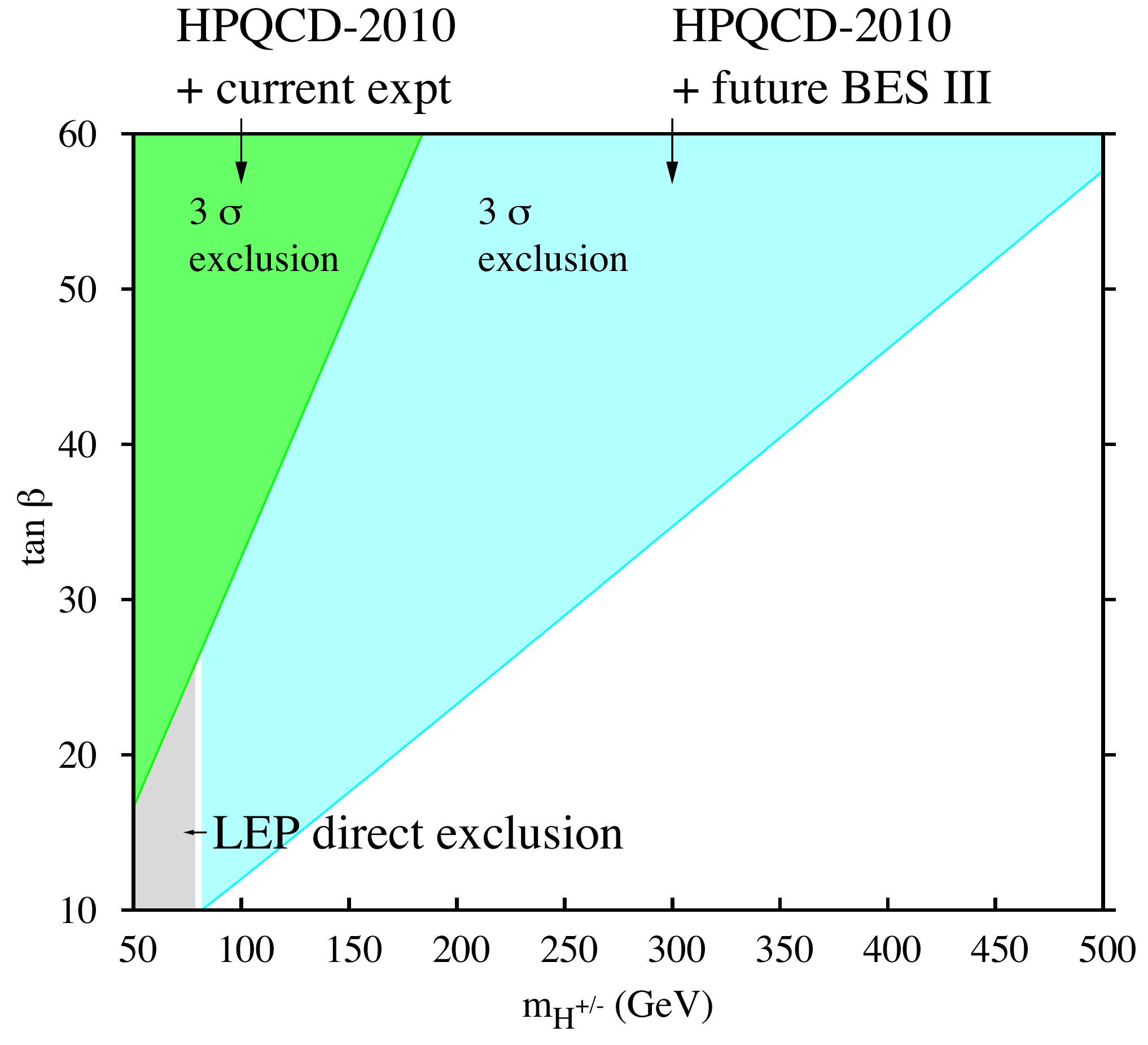}
\end{center}
\caption{ Bounds in the $\tan\beta$/charged Higgs mass plane 
for a 2-Higgs doublet model of Type II given by our 
lattice QCD (i.e. Standard Model) result for $f_{D_s}$
and two different experimental scenarios. The green area 
is excluded at 3$\sigma$ by existing experimental results~\cite{hfag} 
and the light blue area will be excluded by BESIII results~\cite{bes} 
if the central experimental value does not change. 
The light grey band shows the direct limit from LEP searches~\cite{lephiggs}.
}
\label{fig:bsm}
\end{figure}

Instead of assuming a value for $V_{cs}$ to obtain 
an experimental result for $f_{D_s}$ to compare to 
lattice QCD we can combine our result for $f_{D_s}$ 
with the experimental leptonic branching fraction 
to give a direct determination of $V_{cs}$. 
To do this we take the HFAG determination~\cite{hfag} of the 
world average leptonic branching fractions for the 
$D_s$ to $\mu \nu$ and $\tau \nu$ of 0.590(33)\% and 
5.29(28)\% respectively, our result for $f_{D_s}$ and 
\begin{equation}
V_{cs} = \frac{1}{G_F f_{D_s} m_l (1-m_l^2/m_{D_s}^2)}\sqrt{\frac{8\pi{\cal{B}}(D_s \rightarrow l \nu)}{m_{D_s}\tau_{D_s}}}.
\end{equation} 
This gives results for $V_{cs}$ of :
\begin{eqnarray}
V_{cs} &=& 1.033(31), \quad \quad D_s \rightarrow \mu \nu \\ \nonumber
&=& 0.990(28) \quad \quad D_s \rightarrow \tau \nu 
\end{eqnarray}
where the error is dominated by the experimental branching 
fraction. We can combine the results, allowing for correlated 
errors in $f_{D_s}$ and $\tau_{D_s}$, to obtain 
\begin{equation}
V_{cs} = 1.010(22)
\end{equation}
This central value is in a disallowed region 
above 1 so we also provide an alternative 
result that takes this into account. We divide the 
error above into its statistical and systematic 
contributions as 1.010(20)(11) and then reinterpret the 
statistical probability distribution as a Gaussian cut off 
at 1. We then take the central value as the median of 
this new distribution and the error bars as encompassing 
$\pm$ one third of the area about the median.   
This procedure gives the following result: 
\begin{equation}
V_{cs} = 0.990 {+0.007 \atop -0.012} \pm 0.011.
\end{equation}
Both these values for $V_{cs}$ are compatible with 
CKM results (or $V_{ud}$) at better than the 2$\sigma$ level. 
An independent direct determination of 
$V_{cs}$ is possible from $D\rightarrow K l \nu$ semileptonic 
decay for which it is also possible to obtain very 
accurate results with the HISQ action~\cite{hpqcdinprep}. 

A useful bound can be obtained on the mass of a charged 
Higgs from comparing the experimental determination of 
the $D_s$ leptonic branching fraction 
to the expected result using 
$f_{D_s}$ from lattice QCD (i.e Standard Model), 
see, for example,~\cite{akeroyd}.  
In a 2-Higgs doublet model (Type II) the $D_s$ can also 
annihilate to a charged Higgs which interferes destructively 
with the $W$ annihilation. This changes the leptonic 
branching fraction by a simple factor $r$, where 
\begin{equation}
\sqrt{r} = 1 + \frac{1}{1+m_s/m_c}\left(\frac{m_{D_s}}{m_{H^{\pm}}}\right)^2\left(1-\frac{m_s}{m_c}\tan^2\beta\right) 
\label{eq:r} 
\end{equation}
and $\tan\beta$ is the ratio of vacuum expectation values of 
the two scalar doublets. $r < 1$ for large $\tan\beta$ but this 
would be seen from an experimental determination of $f_{D_s}$ 
(using $V_{cs}$ from CKM unitarity)
being smaller than the lattice QCD result. Thus we can derive 
a bound in the $\tan\beta$/$m_{H^{\pm}}$ plane from the 
fact that this is not the case. Here we update what was done 
in~\cite{akeroyd} to include our new lattice QCD result given here and 
the current world average $f_{D_s}$ from experiment~\cite{hfag}. 
These combine to give a central value and error for $\sqrt{r}$ = 1.038(23), 
i.e. $\sqrt{r} > 0.968$ at the $3\sigma$ level. 
Equation~\ref{eq:r}, using our recent accurate determination 
of $m_c/m_s$ from lattice QCD~\cite{mcspaper}, 
then excludes low values of $m_{H^{\pm}}$ as indicated in 
Figure~\ref{fig:bsm}. The bound is not as strong as in~\cite{akeroyd} 
because of the upward shift of our lattice QCD result. However 
the fact that our lattice result, and now the experimental 
average, are so accurate still means that a bound exists. 
New results from BES~\cite{bes} with improved experimental errors would 
produce a much stronger bound, if the experimental central 
value does not change but the error on $f_{D_s}$ is reduced 
to 1\%. This is also indicated in Figure~\ref{fig:bsm}.  
The exclusion limits should be compared to that 
from direct searches at LEP ($m_{H^{\pm}} > $ 78.6 GeV at 95\% C.L.) 
from~\cite{lephiggs} and the estimates of discovery potential 
and exclusion reach of ATLAS at LHC~\cite{atlashiggs}. 
Ref~\cite{ckmfitter} obtains a bound of $m_{H^{\pm}} >$ 316 GeV
from combining results from several processes including $D/D_s$
leptonic decay. 

We have also updated results for $f_{\pi}$, $f_K$ and $f_D$ based 
on the change in the calibration of the lattice spacing used here 
for $f_{D_s}$ but, however, with {\it no} new calculations in 
these cases. We find results consistent with experiment. 
Finally we have given a new very accurate result for $f_{\eta_c}$ 
which will be useful as a calibration point for future 
lattice QCD calculations in charm physics. 

{\bf{Acknowledgements}} We are grateful to the MILC collaboration for the use of 
their configurations.  
Computing was done at the Ohio Supercomputer Center and the Argonne 
Leadership Computing Facility at Argonne National Laboratory, supported
by the Office of Science of the U.S. Department of Energy under 
Contract DOE-AC02-06CH11357. We acknowledge the use of Chroma~\cite{chroma} for part 
of our analysis. 
This work was supported by 
the Scottish Universities Physics Alliance, STFC, MICINN, NSF and DOE.  

\appendix

\section{Sea quark masses}
\label{app:sea}

The staggered quarks in the sea are asqtad improved staggered 
quarks whereas the valence quarks are HISQ quarks, i.e. they use different 
discretisations of the quark piece of the QCD Lagrangian. 
The $s$ quark mass in the two formalisms will then not be the 
same, but there should be a fixed ratio between the two
which is in principle calculable in perturbation theory 
up to discretisation effects. 
This reflects the fact that the difference between the  
two Lagrangians is a difference of regularisation and therefore 
an ultraviolet effect.  Calculations in $\cal{O}$$(\alpha_s)$ 
perturbation theory of the mass renormalisation in the 
two formalisms shows that the $\cal{O}$$(\alpha_s)$ term in 
the relative normalisation is very small~\cite{massren, eigs2}. We therefore 
expect~\cite{massrat} 
\begin{equation}
\frac{am^{hisq}}{am^{asq}} = 1 - 0.004 \alpha_s(a) + C \alpha_s^2(a) + \ldots
\label{eq:mrat}
\end{equation}
up to discretisation and sea quark mass effects. Here 
$am^{hisq}$ and $am^{asq}$ are the lattice valence quark 
masses for the HISQ and asqtad actions respectively that 
give the same meson mass for a particular meson on a 
given ensemble. Note that $am^{asq}$ is defined in the 
conventional way i.e. without the $u_0$ factor present in 
Table~\ref{tab:params}. 

\begin{figure}
\begin{center}
\includegraphics[width=80mm]{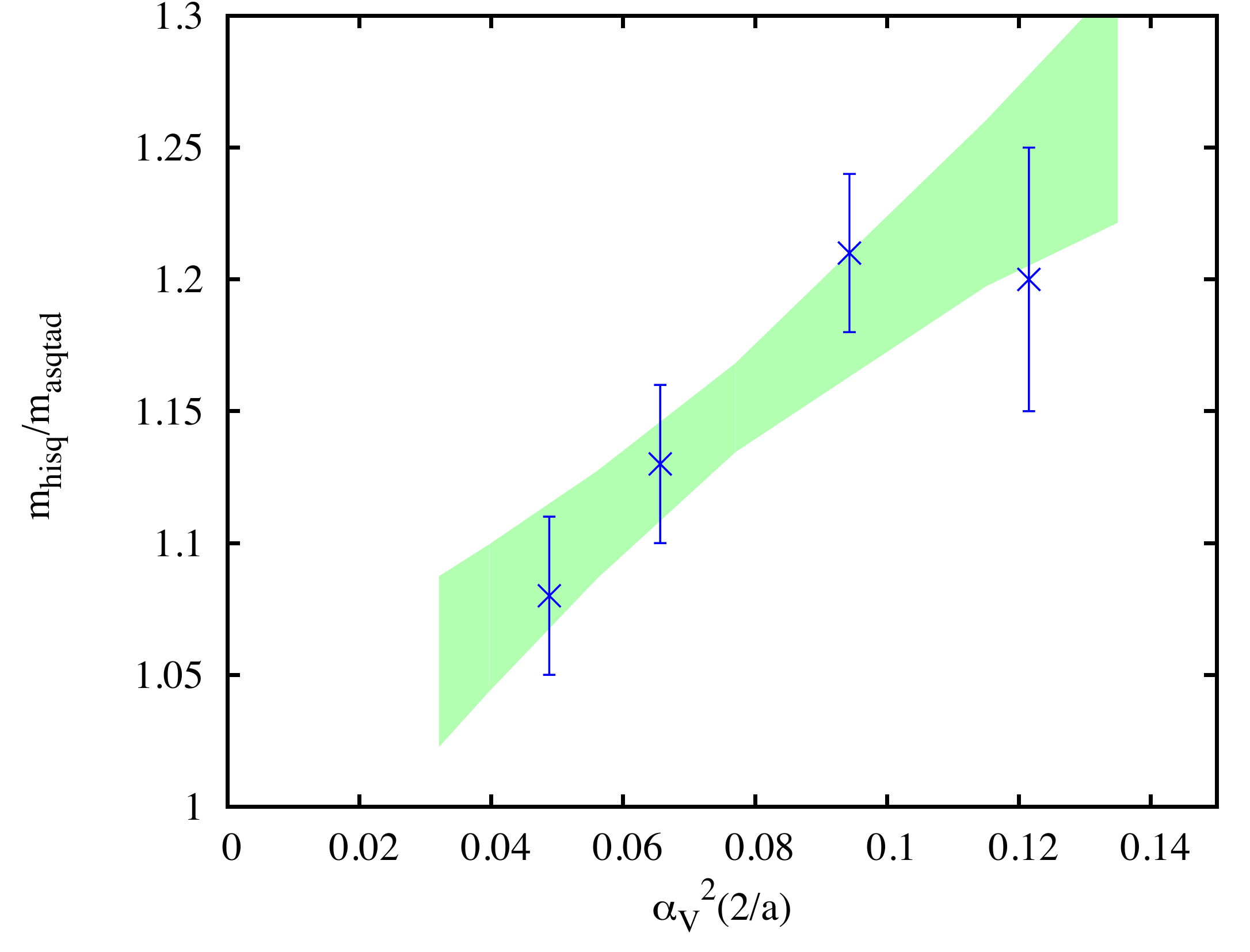}
\end{center}
\caption{Results for the ratio of the physical masses of 
the strange quark using the HISQ formalism~\cite{r1paper} and the 
asqtad formalism~\cite{milcreview}. The mass for the asqtad formalism in 
this ratio has the factor of $u_0$ removed.  
The points are lattice results combining our work and that of 
the MILC collaboration. The shaded band represents a fit of 
the form given in equation~\ref{eq:fitmrat}. 
 }
\label{fig:mqsea}
\end{figure}

Given the HISQ to asqtad mass ratio we can determine 
the tuning of the sea quark masses from our tuning 
of the valence HISQ masses. 
There is very little sea quark mass dependence in the 
quantities that we study here, so that we do not need to 
know this ratio accurately. In principle 
it could be done very accurately, because as we have seen 
the meson masses can be determined very accurately. 
In the absence of this information for asqtad quarks, however, 
we take the suggested tuned asqtad strange quark masses 
from the MILC collaboration~\cite{milcreview}, 
correcting for the $u_0$ factor (taken from the lightest
sea quark mass ensemble at each lattice spacing and 
given in Table~\ref{tab:delta}), and 
compare them to our tuned HISQ strange quark masses~\cite{r1paper}. 
Figure~\ref{fig:mqsea}  shows results on very coarse, coarse, fine and 
superfine lattices. The errors on each point are substantial, 
$\sim$ 3\%, because we have included the tuning error from 
each action added in quadrature, since the tunings were done in a 
different way.  The results can easily be fit to the 
form: 
\begin{equation}
\frac{am^{hisq}}{am^{asq}} = 1 - 0.004 \alpha_s(a) + C \alpha_s^2(a) + Da^2 + Ea^4
\label{eq:fitmrat}
\end{equation}
adding discretisation errors to that in equation~\ref{eq:mrat}.
The fit gives a coefficient $C \approx 2$.   

\begin{table}
\caption{\label{tab:delta}
The sea asqtad masses given in table~\ref{tab:params} have a factor 
of $u_0$ equal to the fourth root of the average plaquette included 
in them. We remove this factor in our comparison of quark masses 
between HISQ and asqtad
and so give values here in column 2 from~\cite{ouralphas}, for all ensembles 
except set 7 where the result is simply estimated from that of the
other coarse lattices.  
Column 3 gives values for the physical asqtad strange quark  at each lattice 
spacing
mass quoted by MILC~\cite{milcreview} and including the $u_0$ factor. 
The result on set 11 is obtained from the tuned HISQ strange mass 
and the ratio described in the text.  
Columns 4 and 5 then give values for $\delta x_l$ and $\delta x_s$ 
as defined in equation~\ref{eq:delta} 
and used in our extrapolations to the physical point. Errors come 
from the errors in $u_0am_{s,phys}$ and are correlated 
between ensembles at a given lattice spacing and between 
$\delta x_l$ and $\delta x_s$ }
\begin{ruledtabular}
\begin{tabular}{lllll}
Set & $u_0$ & $u_0am_{s,phys}^\mathrm{asq}$ & $\delta x_l$ &  $\delta x_s$ \\
\hline
1 & 0.8604  & 0.0439(18) & 0.184(10) & 0.10(5) \\
2 &  0.8610 & 0.0439(18) & 0.405(19) & 0.10(5) \\
\hline
3 &  0.8678 & 0.0350(7) & 0.106(3) & 0.429(29)   \\
4 &  0.8677 & 0.0350(7) & 0.249(6) & 0.429(29) \\
5 &  0.8677 & 0.0350(7) & 0.249(6) & 0.429(29) \\
6 &  0.8688 & 0.0350(7) & 0.535(12) & 0.429(29) \\
7 & 0.868  & 0.0350(7) & 0.249(6) &  -0.143(18) \\
\hline
8 &  0.8782 & 0.0261(5) & 0.201(5) & 0.188(23) \\
9 &  0.8788 & 0.0261(5) & 0.439(10) & 0.188(23) \\
\hline 
10 &  0.8879 & 0.0186(4) & 0.157(5) & -0.03(2) \\
\hline
11 &  0.8951 & 0.0135(5) & 0.170(8) &  0.04(4) \\
\end{tabular}
\end{ruledtabular}
\end{table}

The fitted curve enables us to determine that the 
sea strange quark mass on the ultrafine lattices should 
be 0.0135(5) (with $u_0$ factor {\it included}) i.e it is reasonably 
well-tuned. The error 
is substantial, but the sea quark masses have very little impact on the 
accuracy of results given here. 

Table~\ref{tab:delta} gives values for the $u_0$ parameter (=$(plaq)^{1/4}$)
and the physical asqtad strange quark masses given by the MILC collaboration~\cite{milcreview} for all lattice spacing values except the ultrafine. 
We use our result for ultrafine as discussed above. 
From the physical strange quark mass we determine the physical light quark mass using 
the MILC result: $m_s/m_l = 27.2(3)$.  The table then 
gives values for $\delta x_l$ and $\delta x_s$, where  
\begin{equation}
\delta x_q = \frac{m_{q,sea}-m_{q,sea,phys}}{m_{s,sea,phys}},
\label{eq:delta}
\end{equation}
used in our extrapolation to physical quark masses in section~\ref{sec:res}.


\begin{thebibliography}{99}
\bibitem{footem} 
Note that experimental results 
also include electromagnetic radiation effects 
that are specifically excluded from the definition of $f_{D_s}$ 
and must be subtracted before a value is derived from experiment. 
See the review by J. Rosner and S. Stone in~\cite{pdg}.
\bibitem{hfagnote} Note that the Heavy Flavor Averaging Group adjusts the 
result from~\cite{babar07fds} to account for the fact that it was 
normalised to $D_s \rightarrow \phi \pi$ and this latter branching fraction 
has to be corrected for the mass window applied for $\phi$ reconstruction.
\bibitem{massrat} In~\cite{r1paper} we used a different, more 
phenomenological, form for the HISQ to asqtad mass ratio using 
results from matching meson masses on coarse and fine lattices.
\end{thebibliography}
\end{document}